\documentclass[preprint2]{aastex}
\usepackage{color}
\usepackage{amsmath}
\usepackage{relsize}

\newcommand{\sunrise}{\textsc{Sunrise}}
\newcommand{\figfolder}{./}
\slugcomment{Revised Draft}
\shorttitle{Sunrise II}
\shortauthors{Wiegelmann et al.}
\begin{document}
\title{Magneto-static modelling from \sunrise{}/IMaX: Application
to an active region observed with \sunrise{} II.}
\author{\textsc{
T.~Wiegelmann,$^{1}$
 T.~Neukirch,$^{2}$
 D.H.~Nickeler,$^{3}$
S.~K.~Solanki,$^{1,4}$
P.~Barthol,$^{1}$
A.~Gandorfer,$^{1}$
L.~Gizon,$^{1,9}$
J.~Hirzberger,$^{1}$
T.~L.~Riethm\"uller,$^{1}$
M.~van~Noort,$^{1}$
J.~Blanco~Rodr\'{\i}guez,$^{5}$
J.~C.~Del~Toro~Iniesta,$^{6}$
D.~Orozco~Su\'arez,$^{6}$
W.~Schmidt,$^{7}$
V.~Mart\'{\i}nez Pillet,$^{8}$
\& M.~Kn\"olker,$^{10}$
}}
\date{Draft version \today .}
\affil{
$^{1}$Max-Planck-Institut f\"ur Sonnensystemforschung, Justus-von-Liebig-Weg 3,
37077 G\"ottingen, Germany\\
$^{2}$School of Mathematics and Statistics, University of St. Andrews,
St. Andrews KY16 9SS, United Kingdom \\
$^{3}$Astronomical Institute, AV CR, Fricova 298,
25165 Ondrejov, Czech Republic \\
$^{4}$School of Space Research, Kyung Hee University, Yongin, Gyeonggi, 446-701, Republic of Korea\\
$^{5}$Grupo de Astronom\'{\i}a y Ciencias del Espacio, Universidad de Valencia, 46980 Paterna, Valencia, Spain\\
$^{6}$Instituto de Astrof\'{\i}sica de Andaluc\'{\i}a (CSIC), Apartado de Correos 3004, 18080 Granada, Spain\\
$^{7}$Kiepenheuer-Institut f\"ur Sonnenphysik, Sch\"oneckstr. 6, 79104 Freiburg, Germany\\
$^{8}$National Solar Observatory, 3665 Discovery Drive, Boulder, CO 80303, USA\\
$^{9}$Institut f\"ur Astrophysik, Georg-August-Universit\"at G\"ottingen,
Friedrich-Hund-Platz 1, 37077 G\"ottingen, Germany\\
$^{10}$High Altitude Observatory, National Center for Atmospheric Research,
\footnote{The National Center for Atmospheric Research is sponsored by
the National Science Foundation.} P.O. Box 3000, Boulder, CO 80307-3000, USA\\
}
\email{wiegelmann@mps.mpg.de}
\begin{abstract}
Magneto-static models may overcome some of the issues facing force-free
magnetic field extrapolations. So far they have seen limited use and have faced
problems when applied to quiet-Sun data. Here we present a first application
to an active region.
We use solar vector magnetic field measurements gathered by
the IMaX polarimeter during the flight of the  \sunrise{} balloon-borne
solar observatory in June 2013 as boundary
condition for a magneto-static model of the higher solar atmosphere
above an active region. The IMaX data are embedded
in active region vector magnetograms observed with SDO/HMI. This work
continues our magneto-static extrapolation approach, which has been applied
earlier ({\it Paper I}) to a quiet Sun region observed with \sunrise{} I.
In an active region the signal-to-noise-ratio in the measured Stokes
parameters is considerably higher than in the quiet Sun and
consequently the IMaX measurements
of the horizontal photospheric magnetic field  allow us to specify the free
parameters of the model in a special class of linear magneto-static equilibria.
The high spatial resolution of IMaX (110-130 km, pixel size 40 km)
enables us to model the non-force-free layer between the photosphere and
the mid chromosphere vertically by about 50 grid points.
 In our approach we can incorporate some aspects of the mixed beta layer
of photosphere and chromosphere, e.g., taking a finite Lorentz force into
account, which was not possible with lower resolution photospheric
measurements in the past. The linear model does not, however, permit
to model intrinsic nonlinear structures like strongly localized electric
currents.
\end{abstract}
\keywords{Sun: magnetic topology---Sun: chromosphere---Sun: corona---Sun: photosphere}

\section{Introduction}
Getting insights into the structure of the upper solar atmosphere
is a challenging problem, which is addressed observationally
and by modelling \citep[][]{2014A&ARv..22...78W}.
A popular choice for modelling the coronal magnetic field are
so called force-free configurations
\citep[see][for a review]{2012LRSP....9....5W},
because of the low plasma
$\beta$ in the solar corona above active regions,
\citep[see][]{gary_01}. A complication with this approach is that
necessary boundary conditions for force-free modelling,
namely the vector magnetograms, are routinely observed mainly
in the solar photosphere, where the force-free assumption
is unlikely to be valid
\citep[see, e.g.,][for consequences on force-free models.]
{2009ApJ...696.1780D,2015ApJ...811..107D}.

 A principal way
to deal with this problem is to take non-magnetic forces into
account in the lower solar atmosphere (photosphere to mid chromosphere)
and to use
force-free models only above a certain height, say about
2 Mm, where the plasma $\beta$ is sufficiently low.
Due to the insufficient spatial resolution of vector magnetograms in the past
(e.g. pixel size about 350 km for SDO/HMI, which corresponds
to a resolution of 700 km), trying to
include the relatively narrow lower non-force-free layer in a
meaningful way was questionable.
Nevertheless even with the low resolution of SOHO/MDI magnetograms
(pixel size 1400 km), linear magneto-static models have been applied
in a very limited number of cases
(e.g., by \citeauthor{1998A&A...335..309A} \citeyear{1998A&A...335..309A},
\citeyear{1999A&A...342..867A} to model prominences and
\citeauthor{2000A&A...356..735P} \citeyear{2000A&A...356..735P} developed
a Green's function approach, which was applied to coronal structures in
\citeauthor{2000PhDT.........2P} \citeyear{2000PhDT.........2P}.)
Axis-symmetric magneto-static equilibria  have been applied
in \cite{2008ApJ...689.1379K} to model sunspots from the sub-photosphere to
the chromosphere.

The vast majority of active region models are, however, based
on the force-free assumption  \citep[see, e.g.,][for an overview, covering
both linear and non-linear force-free models]{1997SoPh..174..129A}
 and one has to deal with the problem that the photospheric
magnetic field vector has measurement inaccuracies
\citep[see, e.g.,][how these inaccuracies affect the quality of force-free
field models]{2010A&A...511A...4W} and the photosphere
is usually not force-free,
\citep[see, e.g.,][]{1995ApJ...439..474M}.
One possibility to deal with this problem is to apply Grad-Rubin codes, which do
not use the full photospheric field vector as boundary condition, but
the vertical magnetic field $B_z$ and the vertical electric current density
$J_z$. The latter quantity is derived from the horizontal photospheric field.
The Grad-Rubin problem is well posed, if $J_z$ (or alternatively
$\alpha=J_z/B_z$) is prescribed only for one polarity of the magnetic field
and the two solutions ($\alpha$ prescribed for the positive or negative
polarity) can differ significantly \citep[see][]{2008ApJ...675.1637S}.
Advanced Grad-Rubin codes take $J_z$ (or $\alpha$) and measurement
errors on both polarities into account
\citep[see, e.g.][for details]{2009ApJ...700L..88W,2010A&A...522A..52A}.
An alternative approach, dubbed {\it pre-processing}, was introduced in
\cite{2006SoPh..233..215W}
to bypass the problem of inconsistent photospheric vector magnetograms
by applying a number of necessary (but not sufficient) conditions
to prescribe boundary conditions for a force-free modelling.
 Resolving the physics of the thin mixed plasma $\beta$ layer was not aimed at
in this approach and was also not possible due to
observational limitations. The reason is that for meaningful
magneto-static modelling, the thin non-force-free region
(photosphere to mid-chromosphere, about 2 Mm thick) has to be resolved
by a sufficient number of points.

Naturally the vertical resolution
of the magneto-static model scales with the horizontal spatial
scale of the photospheric measurements. With a pixel size of 40 km
for data from \sunrise{}/IMaX, we can model this layer with 50 grid points.
We have applied the approach to a quiet-Sun region measured by
\sunrise{}/IMaX during the 2009-flight in
\cite{2015ApJ...815...10W} ({\it Paper I})
and refer to this work for the mathematical and computational
 details of our  magneto-static
code. Here we apply the method to an active region measured by
\sunrise{}/IMaX during the 2013-flight.
This leads to a number of differences due to the
different nature of quiet and active regions. In active regions we get
reliable measurements of the horizontal photospheric field vector,
which was not the case in the quiet Sun due to the poor signal-to-noise
ratio \citep[see][for details]{2011A&A...527A..29B,2012A&A...547A..89B}.
Dealing with an active region also requires differences in procedure.
While the spatial resolution of IMaX is very high, the FOV is limited to parts
of the observed active region.
For a meaningful modelling one has
to include, however, the entire active region and a quiet-Sun skirt
around it in order to incorporate the magnetic connectivity and
as well the connectivity of the related electric currents.
This requirement on the FOV was originally pointed out
for force-free modelling codes
\citep[][]{2009ApJ...696.1780D}, but remains valid for the magneto-static
approach applied here. Consequently we have to embed the IMaX measurements
into vector magnetograms from SDO/HMI
 \citep[see][for an overview on the SDO mission
and the HMI instrument, respectively.]{2012SoPh..275....3P,2012SoPh..275..207S}.
This was not necessary for the quiet
Sun configurations in {\it Paper I}.

This paper provides the first test of our new method in an active region.
Since active region fields (sunspots, pores) are often stronger than those
in the quiet Sun, it is not a priori clear if and how the method
can be applied to an active region. Our aim is to carry out the corresponding
tests and address the related complications and limitations.
The outline of the paper is as follows:
In Section \ref{sec:data} we describe the used data set from \sunrise{}/IMaX,
which we embed and compare with measurements from SDO/HMI.
The very different resolution of both instruments (almost a factor of ten)
leads to a number of complications, which are pointed out and discussed.
Section  \ref{sec:theory} contains a brief reminder on the used special class
of magneto-static equilibria.  As
the details of the model are described
in {\it Paper I}, we only describe the adjustments we make for active-region
modelling. Different from the quiet Sun, we are able to deduce and specify
all free model parameters from measurements. In Section \ref{sec:results}
we show a few example field lines for two (out of 28 performed) snapshots,
and the related self-consistent plasma properties (plasma pressure
and plasma $\beta$). We point out some differences of magneto-static
equilibria to potential and force-free models. We perform
a statistical analysis of loops, but
a detailed analysis of the magneto-static time series
is outside the scope of this paper.
Finally we discuss the main features and problems of the active region
magneto-static modelling introduced here in section \ref{sec:outlook}.
\section{Data}
\label{sec:data}
\begin{figure*}%[h]
\includegraphics[width=0.95 \textwidth]{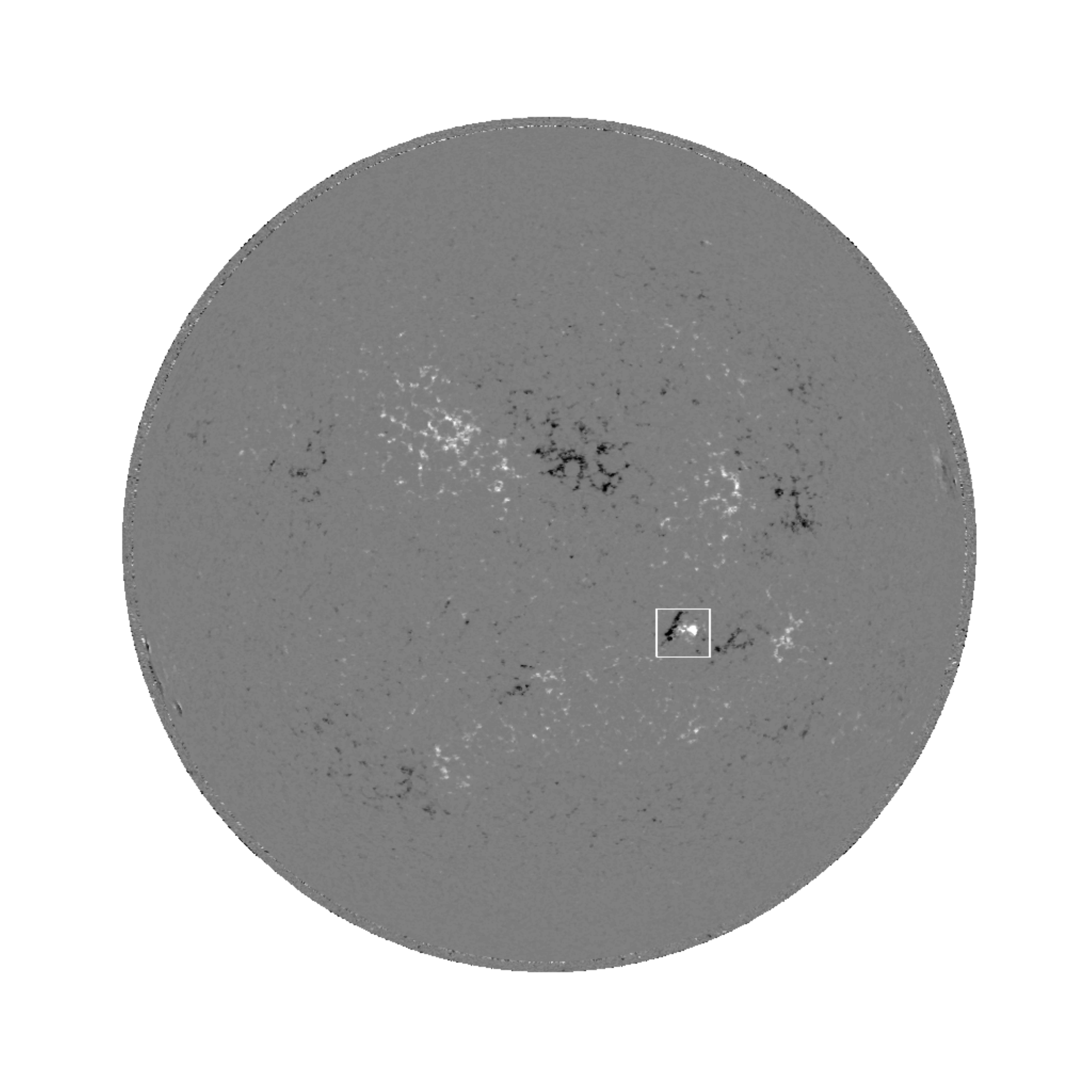}
\caption{The full Sun observed with SDO/HMI  on
2013, June 12th at 23:40UT. The white box marks
the active region  AR11768, investigated in this paper.}
\label{figure_full}
\end{figure*}
The \sunrise{} balloon-borne solar observatory
\citep[see][for details]{2011SoPh..268....1B,2011SoPh..268..103B,2011SoPh..268...35G}
carries the vector magnetograph IMaX \citep[][]{martinezpillet:etal11}.
\sunrise{} has flown twice, the first time in 2009
\citep[][referred to as \sunrise{} I]{solanki:etal10}
when it exclusively observed quiet Sun.
These data were inverted by \cite{borrero:etal11}
using the VFISV code and more recently again,
after further refinements, by \cite{kahil:etal16}
using the SPINOR inversion code \citep{2000A&A...358.1109F}.
\sunrise{} flew again in 2013
(referred to as \sunrise{} II) when it caught an emerging
active region. The changes in the instrumentation, the
flight, data reduction and inversions are described by
\cite{solanki:etal16}. The atmospheric model for the inversion
assumes a hight independent magnetic field vector.
In a forthcoming work we plan to use also an
MHD-assisted Stokes inversion (leading to a 3D solar atmosphere),
as described by \cite{tino:etal16}.

Figure \ref{figure_full} shows a full disk image of the
line-of-sight magnetic field observed with SDO/HMI on
2013, June 12th   at 23:40UT and AR11768 is marked with
a white box. A part of this AR has been observed with \sunrise{} II.
For the work in this paper we use
a data set of 28 IMaX vector magnetograms taken with a cadence of
$36.5$s starting on  2013, June 12th at 23:39UT.
 The data have a pixel size of
40 km and the IMaX-FOV contains $(936 \times 936)$ pixel$^2$
(about $(37 \times 37)$ Mm$^2$).
Due to the high spatial resolution of \sunrise{}/IMaX
and a correspondingly small FOV, we embed the data in vector magnetograms
observed with SDO/HMI at 23:36UT,  23:48UT and at 00:00UT. The combined data set
contains the entire active region, is approximately flux balanced and
the total FOV is $(89 \times 86)$  Mm$^2$. The location of the active region
is marked with a white box in Fig. \ref{figure_full}.

\subsection{Embedding and ambiguity removal}
\begin{figure*}%[h]
\includegraphics[width=0.95 \textwidth]{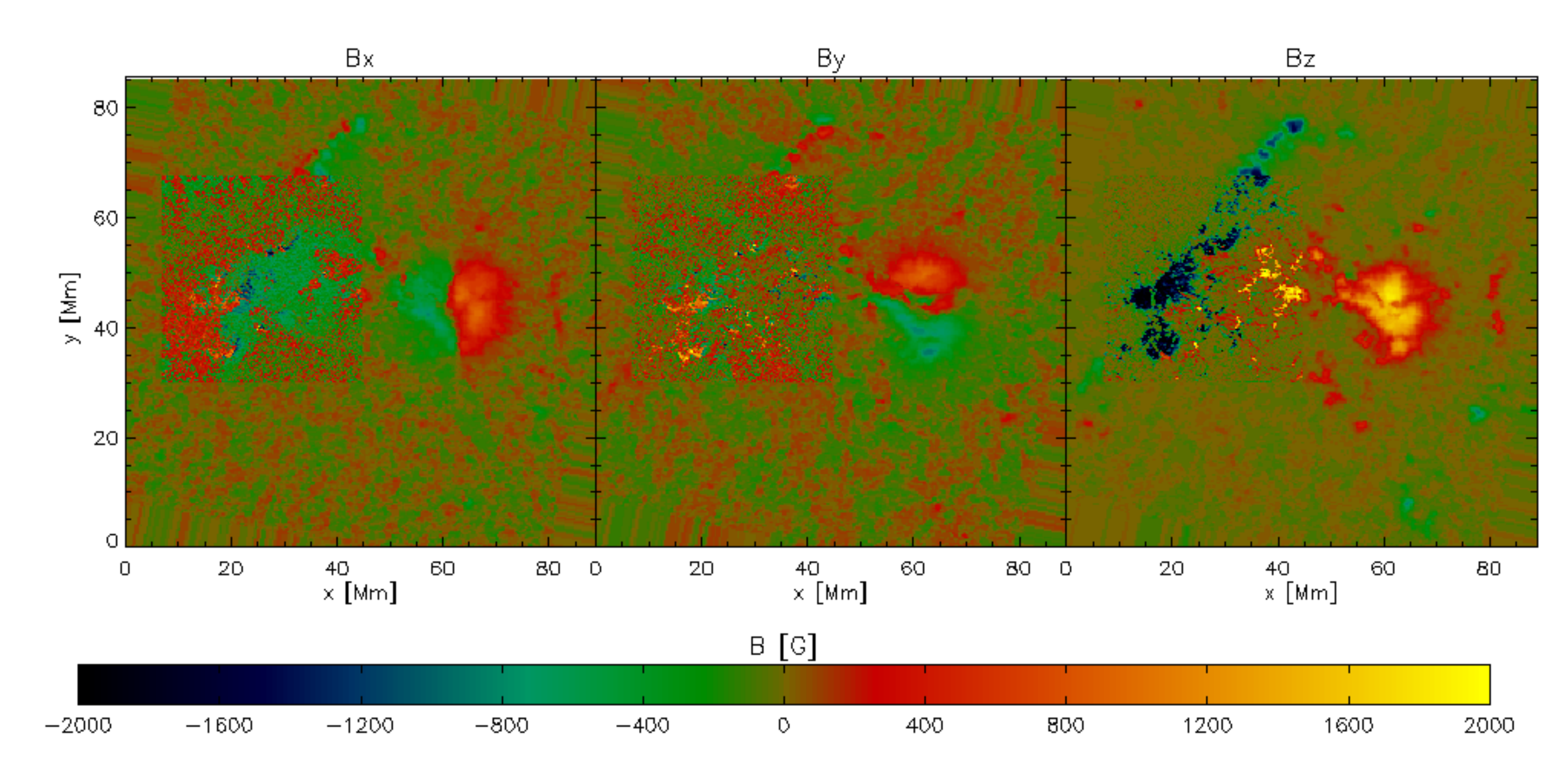}
\includegraphics[width=0.95 \textwidth]{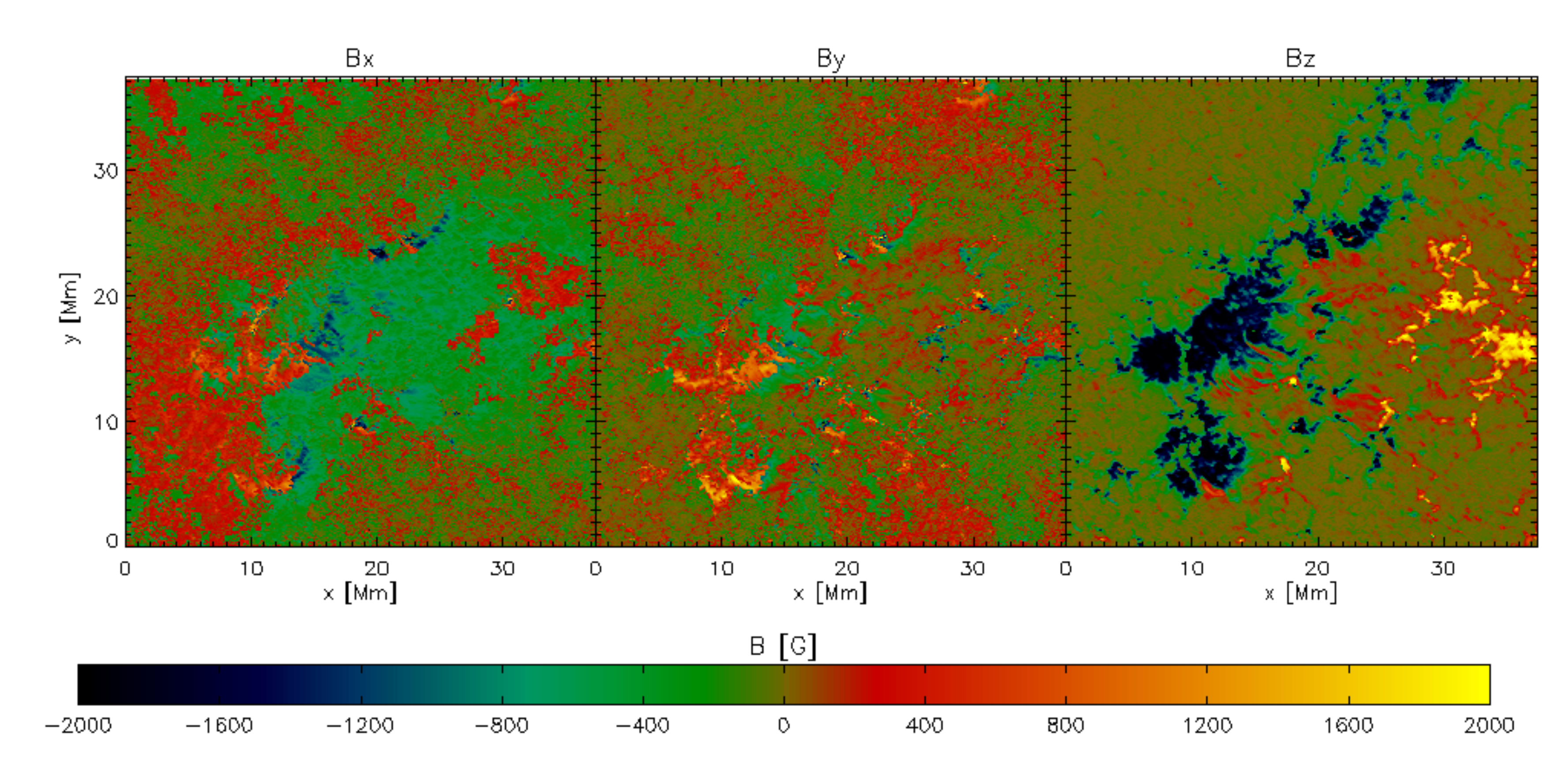}
\caption{Top: Vector magnetogram of IMaX (first snapshot
taken on 2013, June 12th at 23:39UT)
embedded in the HMI FOV. The FOV of IMaX is clearly visible
due to the better resolved structures. The left and center panels
show the horizontal field components $B_x$ and $B_y$.
The right panel corresponds to the vertical field $B_z$.
Bottom: vector magnetogram for the IMaX FOV. Please note the different
x- and y-axes in top and bottom panels.}
\label{vecmag}
\end{figure*}

\begin{figure*}
\mbox{
\includegraphics[width=8.0cm, height=6.7cm]{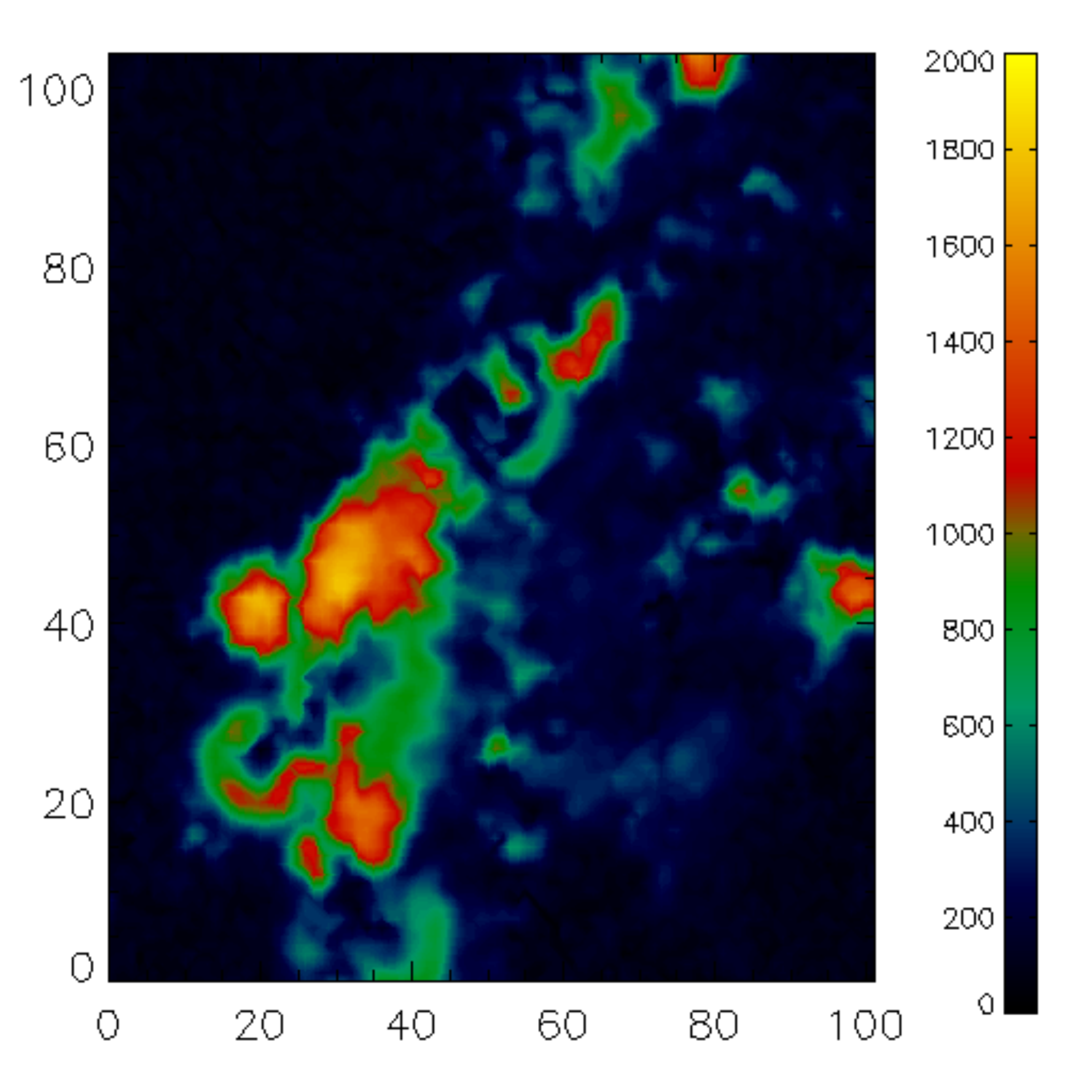}
\includegraphics[width=8.0cm, height=6.7cm]{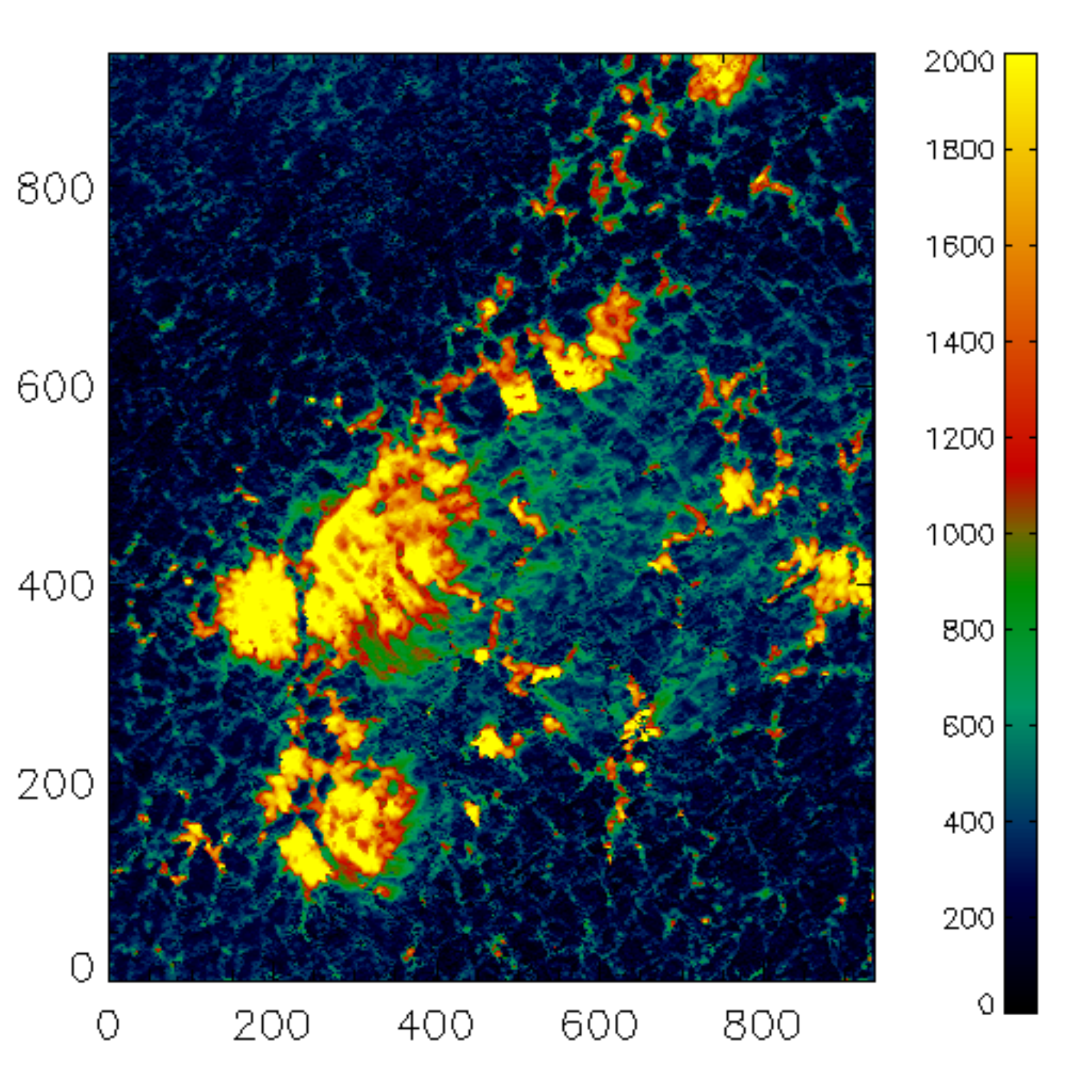}}
%}
\caption{Comparison of the field strength $B$ (only within the FOV of IMaX) for
data from HMI (left) and IMaX (right).
The x- and y-axes are numbered in pixels.
Due to the much higher resolution of IMaX, stronger fields are
detected.  Naturally
this results also in a higher average field strength in
the IMaX data: $470 G$ than those from
HMI: $287 G$. Both data-sets have been taken almost at
the same time at 23:48UT.}
\label{compare_hmi_imax}
\end{figure*}

To align the HMI vector maps and IMaX vector magnetogramms
we rotate
($\phi \approx -10^{\circ}$) and rescale (by about a factor 9) the
HMI-data. The exact values are computed separately
for each snapshot by a correlation analysis.
From the three HMI vector magnetograms we always choose the one closest
in time to the related IMaX snapshot.
The horizontal field vectors from HMI are transformed by the rotation
to the local coordinates of the IMaX-FOV
\citep[see][for the transformation procedure]{1990SoPh..126...21G}.
We note that this effect is very small for the small rotation angle
of $\phi \approx -10^{\circ}$ found here. The correlation between fields with
and without taking this effect into account is $98 \%$.
We remove the $180^\circ$ ambiguity
in the IMaX data with an acute angle method. See, e.g.,
\cite{2006SoPh..237..267M} for an overview of ambiguity removal methods.
The acute angle method minimizes the angle with a reference field,
here the corresponding HMI vector magnetograms.
The resulting field is shown in Figure \ref{vecmag}.
On average $13 \% \pm 4\%$ of the pixels flip
their ambiguity between consecutive snapshots.
We note that the chosen HMI magnetograms are almost flux balanced,
with an imbalance of $-0.5 \%, -1.2 \%, -0.5 \%$ respectively.
The combined data set (IMaX embedded in HMI) shows an imbalance
of $-4.2 \% \pm -0.5 \%$. This is a systematic effect which necessarily
appears due to the much higher resolution of IMaX. The net flux is negative,
because the FOV of IMaX is located in a  mainly negative polarity region.
HMI misses a significant amount of small scale magnetic flux, as shown
in Fig. \ref{compare_hmi_imax} (see also the paper by
\cite{chitta:etal16}, who also show that HMI misses a
considerable amount of small-scale flux and structure).
This difference in the flux measured by the two instruments is a natural
result of their different spatial resolutions.
 The missing small scale flux is due to a cancellation of the
Zeeman signals of opposite polarity fields within a resolution
element of HMI. The field strength in HMI-magnetograms is lower,
because the HMI inversion does not use filling factors.

\section{Theory}
\label{sec:theory}
\subsection{Magneto-static extrapolation techniques}
We use the photospheric vector magnetograms described above as
boundary condition for a magneto-static field extrapolation.
Therefore we use a special class of separable magneto-static solutions
proposed by \cite{low91}. This model has the advantage of leading to
linear equations, which can be solved effectively by a fast Fourier
transformation. A corresponding code has been described and applied
to a quiet-Sun region observed with \sunrise{} I in
{\it Paper I}. Here we only briefly describe the main
features of this method and refer to our {\it Paper I} for details.
The electric current density is described as
\begin{equation}
\nabla \times {\bf B } = \alpha {\bf B } + a \exp(-\kappa z) \nabla B_z \times {\bf e_z},
\label{def_j}
\end{equation}
where $\alpha$ controls the field aligned currents and
$a$ the non-magnetic forces, which compensate the Lorentz-force.
Because the solar corona above active regions is almost force-free
\citep[see][]{gary_01} the non-magnetic forces
have to decrease with height.
As in {\it Paper I} we
choose $\frac{1}{\kappa} = 2$ Mm to define the height of
the non-force-free domain.

\subsection{Using observations to optimize the parameters $\alpha$ and $a$}
\label{compute_alpha_a}
\begin{figure*}
\includegraphics[width=0.95\textwidth]{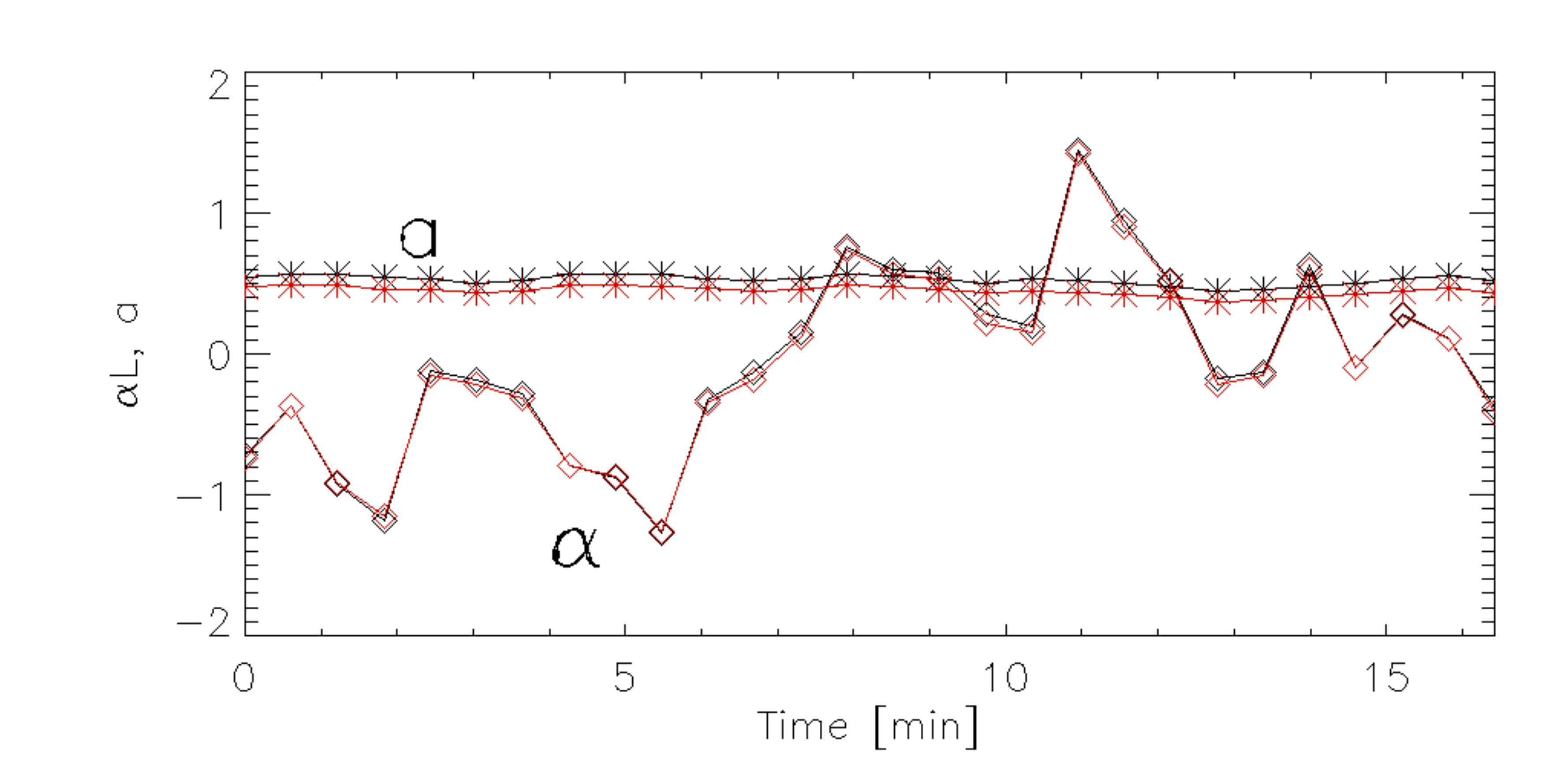}
\caption{
Temporal evolution of $\alpha L$ (diamonds)  and $a$ (asterisks) as
computed by equations (\ref{eq_alpha}) and (\ref{eq_a}).
In black are shown the values computed from the original
IMaX vector magnetograms and in red a re-evaluation from
the resulting magneto-static equilibria. $t=0$ corresponds to 23:39UT.}
\label{time_a_alpha}
\end{figure*}
In {\it Paper I}, $\alpha$ and $a$ were treated as free parameters. For the
active region measurements in this paper, we propose to use the horizontal
photospheric field vector to constrain $\alpha$ and $a$. This was
not possible for the quiet-Sun (QS) region investigated in {\it Paper I}, because
the poor signal-to-noise-ratio in QS regions
does not allow an accurate determination of the horizontal field components.
For computing $\alpha$ we follow an approach developed by
\cite{2004PASJ...56..831H} for linear force-free fields:
\begin{equation}
\alpha=\frac{\mathlarger{\sum} {\left(\frac{\partial B_y}{\partial x}-\frac{\partial
B_x}{\partial y}\right) {\rm sign}(B_z)}}{ \sum{|B_z|}},
\label{eq_alpha}
\end{equation}
where the summation is done over all pixels of the magnetogram.
Please note that $\alpha$ has the dimension of an inverse length
and the values of $\alpha$ presented in this paper are normalized
with $L=37$ Mm, which is the width of the IMaX-FOV.
The temporal evolution of $\alpha$ as
deduced from Eq. (\ref{eq_alpha}) is shown with diamonds
in Figure \ref{time_a_alpha}.
The input (black diamonds) in Fig. \ref{time_a_alpha}
and output values (red diamonds) of the global parameter $\alpha$ are
almost identical. The
small discrepancies that occur are due to numerical errors.

A straightforward way of computing the force parameter $a$ in
Eq. (\ref{def_j}) is more challenging than computing $\alpha$.
While $\alpha$ controls currents which are strictly parallel to the
field lines, this is different for the $a$ term. This part
controls the horizontal currents and in the generic case these currents
are oblique to the magnetic field. This means they have a
parallel as well as a perpendicular component. The latter one
is responsible for a non-vanishing Lorentz force. We recall
\citep[see][for details]{1969SvA....12..585M,1974SoPh...39..393M}
that the Lorentz force can be written as the volume integral of
the divergence of the Maxwell
stress tensor $T$:
\begin{equation}
F_{\rm Lorentz} = \int \nabla \cdot T \,  dV = \oint T \,  ds,
\end{equation}
and one gets  surface integrals enclosing the volume by
applying Gauss' law. This approach
is used frequently in nonlinear force-free computations to
check whether a magnetogram is consistent with the force-free criterion.
In principle the surface integral has to be taken over the entire
surface of the computational volume, but for applications to
measurements, one has to restrict it to the bottom, photospheric boundary.
We note that neglecting the contribution of lateral boundaries can
be more critical for small FOVs like ours than for full ARs surrounded by a
weak field skirt. Following a suggestion by
\cite{1989SoPh..120...19A}, the components of the surface integral (limited to the
photosphere) are combined and normalized to define a dimensionless
parameter:
\begin{equation}
\epsilon_{{\rm force}} =  \frac{\left|\mathlarger{\sum} B_x B_z \right| +
\left|\mathlarger{\sum} B_y B_z \right|+
\left|\mathlarger{\sum} (B_x^2+B_y^2)-B_z^2  \right|}
{\mathlarger{\sum} (B_x^2+B_y^2+B_z^2) },
\label{eq_eps_force}
\end{equation}
where the summation is done over all pixels of the magnetogram.
This parameter is frequently used to check whether a given
vector magnetogram is force-free consistent (and can be used
as boundary condition for a force-free coronal magnetic field modelling)
or if a pre-processing is necessary
\citep[see][for details]{2006SoPh..233..215W,2008SoPh..247..249W}.
While the pre-processing aims  at finding suitable boundary conditions
for force-free modelling, the magneto-static approach used here
takes the non-magnetic forces into account. While $a$ in
Eq. (\ref{def_j}) controls the corresponding parts of the current and
Eq. (\ref{eq_eps_force}) is a measure for the non-vanishing Lorentz-force,
it is natural to try to relate $a$ and $\epsilon_{{\rm force}}$.
Because $a$ is linear in the electric current density, it
implicitly also influences the magnetic field, and we cannot
assume that the relationship of $a$ and $\epsilon_{{\rm force}}$
is strictly linear. Nevertheless an empirical approach suggests
 a linear relation to lowest order and one finds that
\begin{equation}
a = 2 \epsilon_{{\rm force}}
\label{eq_a}
\end{equation}
is a reasonable approximation for specifying the free parameter
$a$. The black asterisks in Fig. \ref{time_a_alpha} show the temporal
evolution of $a$ (or $2 \epsilon_{{\rm force}}$) as deduced with
Eqs. (\ref{eq_eps_force}) and (\ref{eq_a}) from IMaX. We re-evaluate
the forces in the photosphere from the resulting magneto-static equilibrium,
shown as red asterisks in figure \ref{time_a_alpha}. It is found that our special
class of linear magneto-static equilibria somewhat
underestimates the forces  $(15 \% \pm 1 \%)$
in the lowest photospheric layer. This effect
occurs with a low scatter for the entire investigated time series.
Possible reasons for this behaviour are the general limitations
of applying Eq. (\ref{eq_eps_force}) to a small FOV, as discussed above.
We note that a linear model cannot be
assumed to reveal local structures
like localized electric currents and horizontal magnetic fields. This
is a property which linear magneto-static fields share with linear force-free
fields. But the linear magneto-static approach allows considering
the non-force-free nature of the lower solar atmosphere as deduced from
measurements from equation (\ref{eq_eps_force}). A linear force-free model
would fulfill (\ref{eq_alpha}) as well, but $\epsilon_{{\rm force}}$ is
zero per definition for force-free models.
\section{Results}
\label{sec:results}
\subsection{3D magnetic field lines}
\begin{figure*}
\mbox{
\includegraphics[width=0.5\textwidth]{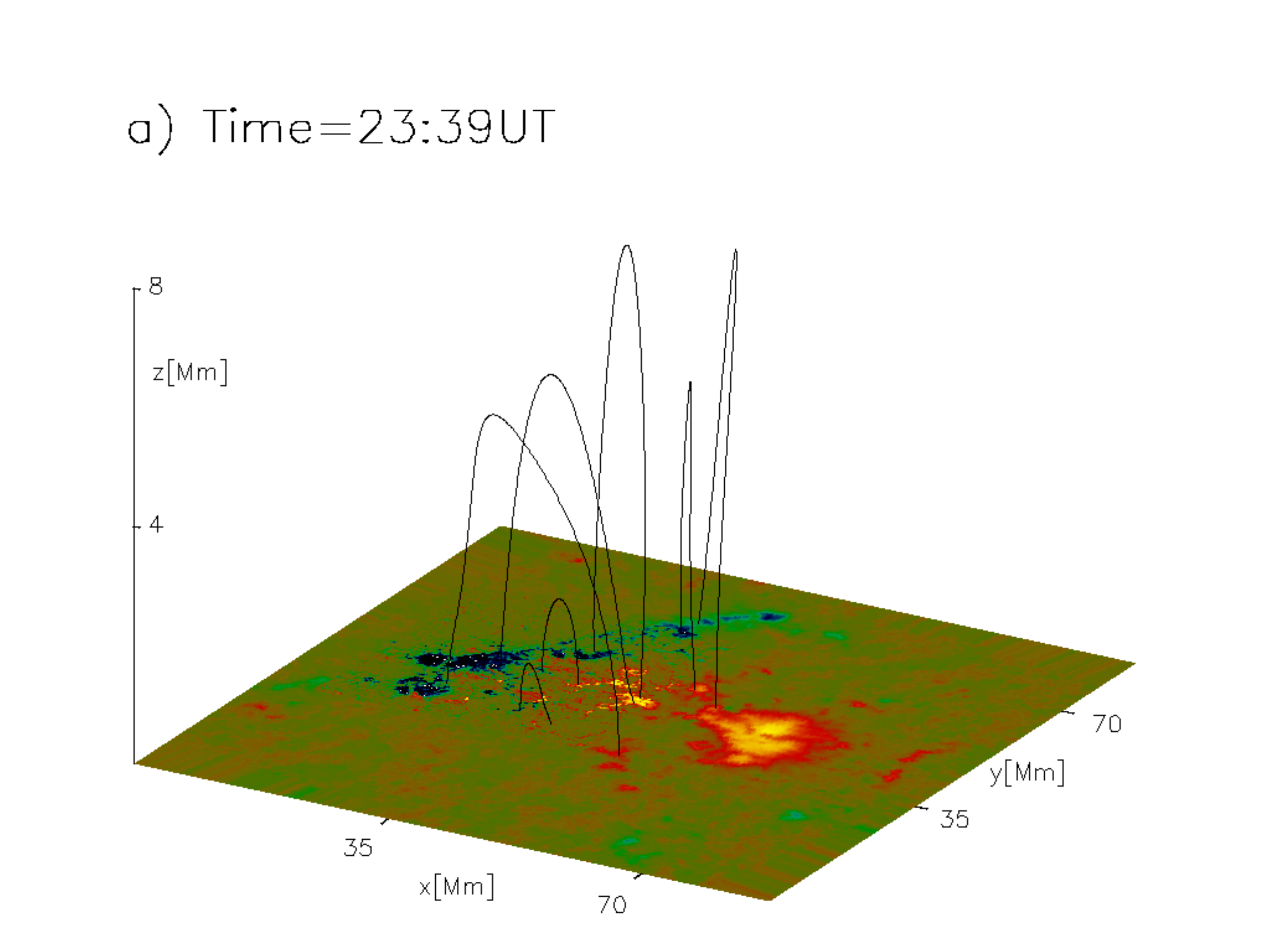}
\includegraphics[width=0.5\textwidth]{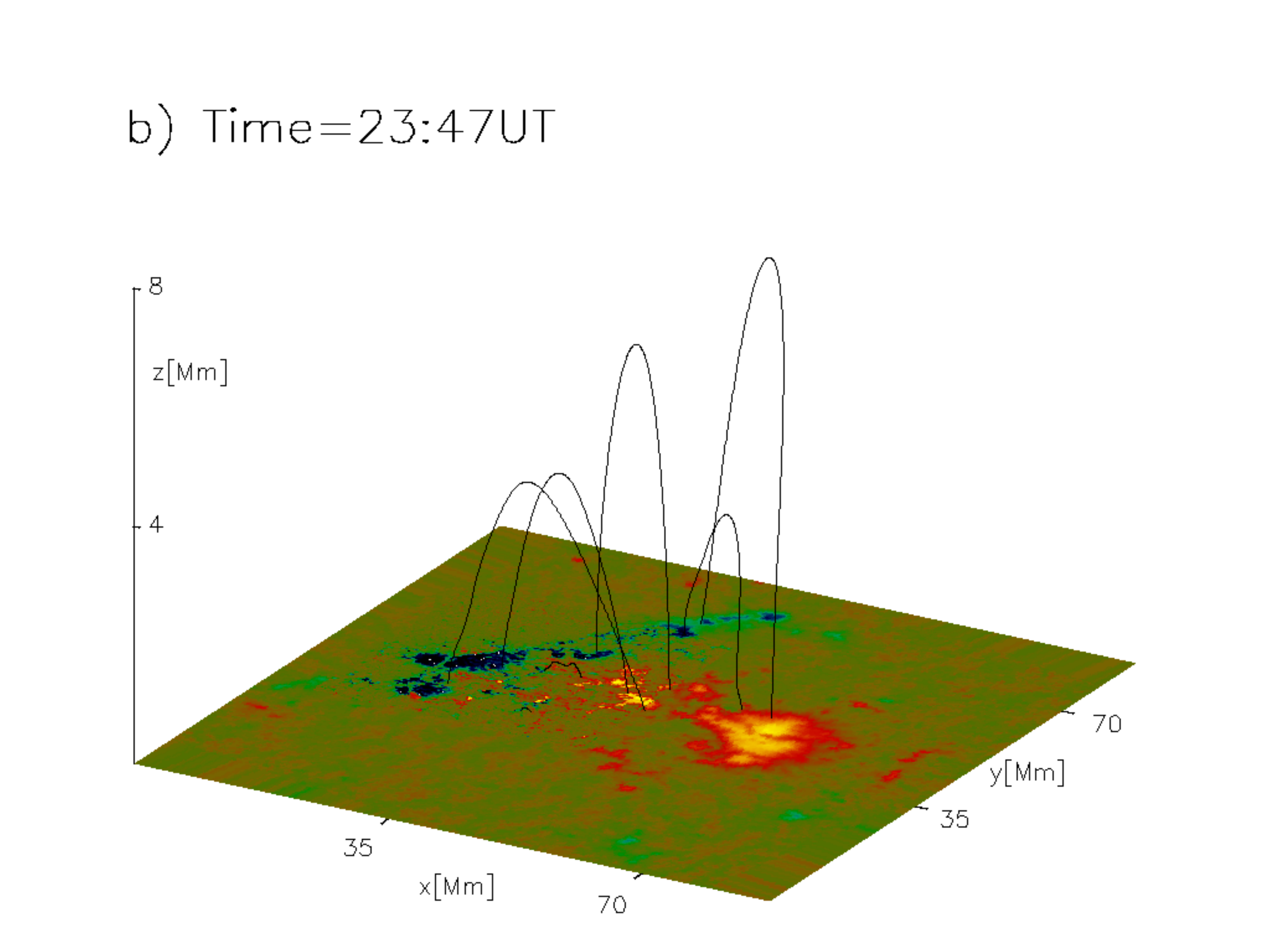}
}
\mbox{
\includegraphics[width=0.5\textwidth]{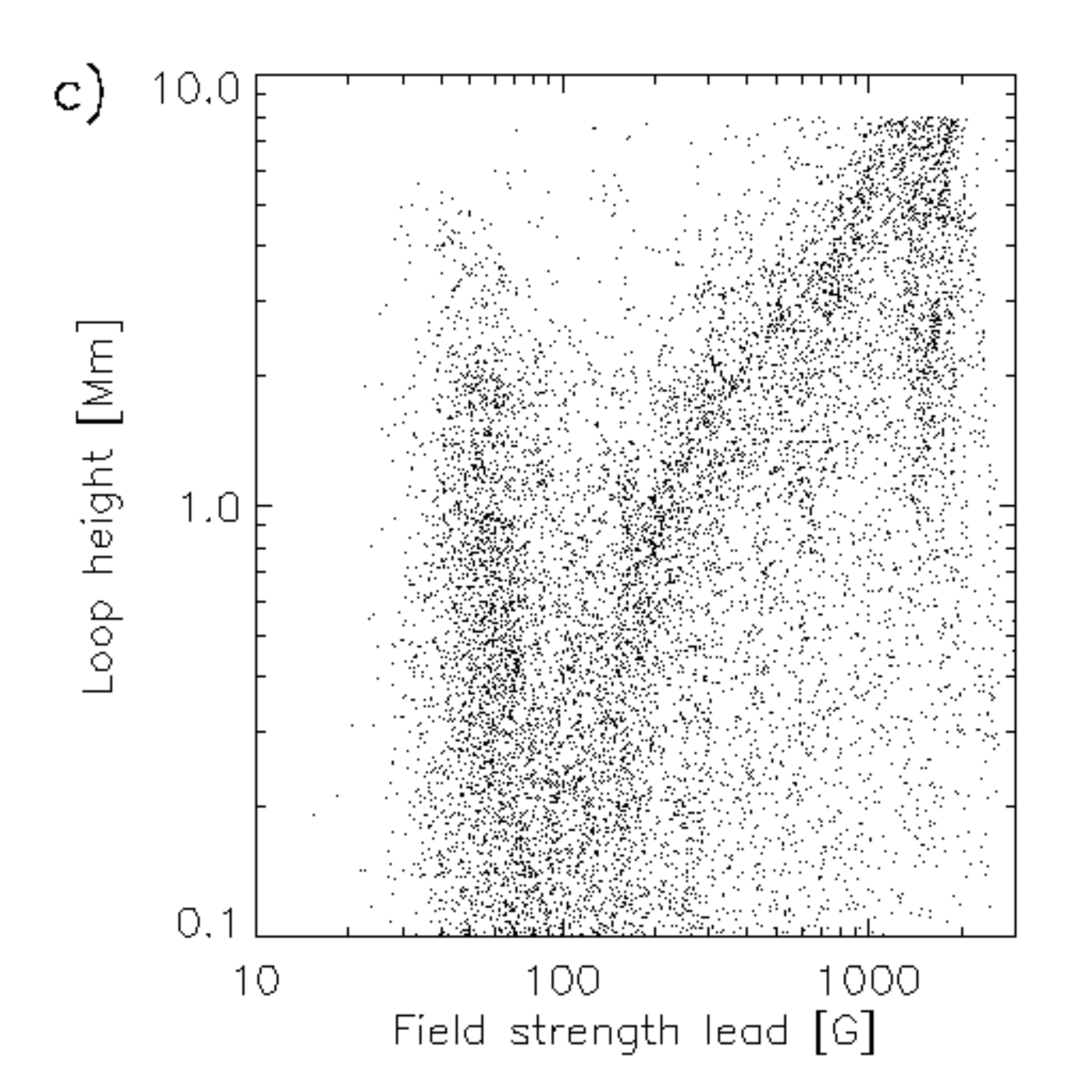}
\includegraphics[width=0.5\textwidth]{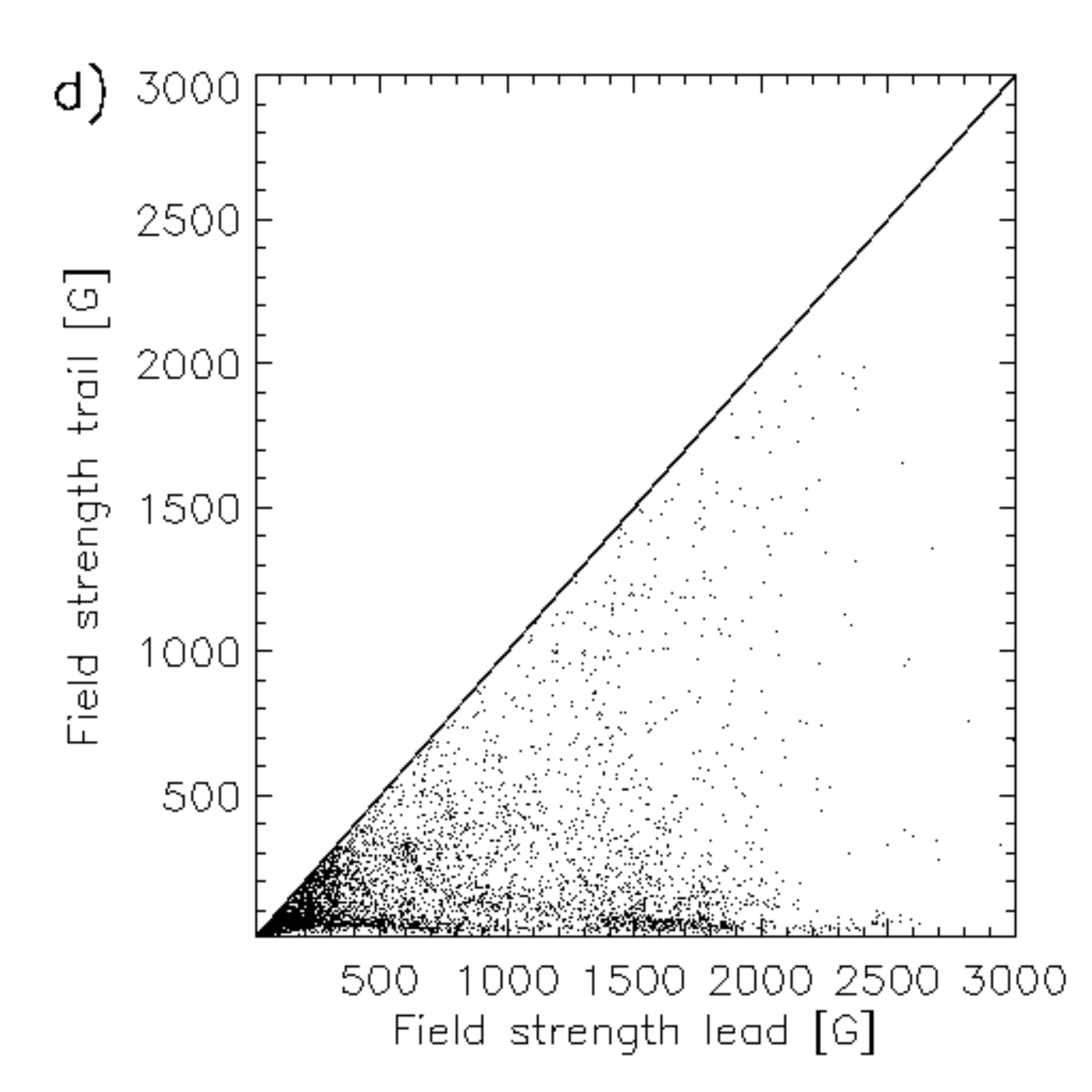}
}
%\mbox{\includegraphics[width=0.5\textwidth]{\figfolder bmhs1}
%\includegraphics[width=0.5\textwidth]{\figfolder bmhs20}}
\caption{Panels a) and b) show  example field lines
for the entire (HMI+IMaX) FOV and up to $z=8$ Mm for the first snapshot
at 23:39UT and at snapshot 15  at 23:47UT.
In panel c) we show for a sample of $10.000$ randomly
chosen loops a scatter plot of the strength at the leading foot points
and the loop heights at 23:39UT. Panel d) shows a scatter plot
of the field strength of the leading (stronger) and trailing
 (weaker) foot points for loops reaching at least into
the chromosphere.}
\label{fig_mhs}
\end{figure*}
In Fig. \ref{fig_mhs} we show a few sample field lines
at 23:39UT and  23:47UT in panels a) and b), respectively.
The field line integration has been started at the same points in
negative polarity regions. As one can see some of the larger, coronal
loops change their connectivity during this time and connect to
different positive polarity regions in both panels.
In panel a) the two smallest loops reach into the chromosphere. This
is not the case in panel b), where these loops close already
 at photospheric heights.
 A detailed analysis of these features is well outside the
scope of this paper, however. Further investigations of these low
lying structures, also taking \sunrise{}/SUFI data into account can
be found in the paper by \cite{shahin:etal16}.

 In the following we investigate the relation of the strength of
 loop foot points and loop heights. Therefor we analyse a sample of
 10,000 randomly chosen loops, excluding loops originating in
 photospheric regions below the
 the $1 \sigma$ noise level of $13$G \citep[the $3 \sigma$ noise level
 is $40G$,see][]{2016arXiv160905664K}, those originating in
 a frame of 150 pixel at
 the lateral boundaries of the magnetogram, unresolved loops
 (loop top below $z=100$km) and field lines not closing within the
 computational domain.

For the  snapshots at 23:39 UT (23:47 UT)
we found a correlation of the stronger, leading foot point strength and loop
height of $51 \% (55 \%) $ and a correlation of the weaker trailing foot point
strength and loop height of $32 \% (40 \%)$.
In Fig. \ref{fig_mhs}c) we show a scatter plot (based on 10,000 loops)
of the strength of the
leading  foot point and height of the loops. In Table \ref{table1},
deduced from two snapshots at 23:39UT and 23:47UT, we investigate
some properties of photospheric, chromospheric and coronal loops.
The values hardly change for a larger sample of loops and
temporal changes are moderate.
\begin{table}[h]
\caption{The table shows the average field strength at the
leading (stronger)  and trailing (weaker) foot points
for loops reaching into different regions of the solar atmosphere.
The first row contains all 10,000 loops,
the second row loops closing within photospheric heights
(loop top below $z <0.5$Mm), the third row chromospheric loops
($0.5 \leq z < 2$ Mm) and the fourth row coronal loops ($z \geq 2$Mm).
The upper part of the table has been deduced from a snapshot at
23:39UT and the lower part from a snapshot at 23:47UT.}
\label{table1}
\begin{tabular}{llccc}
\hline
Time & Region & Perc. & lead [G] & trail [G] \\
\hline
23:39UT & all    &  $100 \%$  & 476 & 152 \\
23:39UT & photosphere  &  $ 39 \%$  & 266 & 114 \\
23:39UT & chromosphere  &  $ 37 \%$ & 395 & 120 \\
23:39UT & corona &  $ 24 \%$  & 947 & 263 \\
\\
23:47UT & all    &  $100 \%$  & 440 & 147 \\
23:47UT & photosphere &  $ 42 \%$  & 238 & 107 \\
23:47UT & chromosphere &  $ 39 \%$ & 380 & 114\\
23:47UT & corona &  $ 19 \%$  & 995 & 297 \\
\hline
\end{tabular}
\end{table}

There is a clear tendency that on average both footpoints of
coronal loops are stronger than for loops reaching only into the
chromosphere or photosphere. While on average the leading footpoint
of chromospheric loops is about a factor of $1.5$ stronger than for photospheric
loops, one hardly finds a difference for trailing footpoints between
photospheric and chromospheric loops.

Fig. \ref{fig_mhs}d) shows a scatter plot of the
leading and trailing foot points for loops reaching at least into
the chromosphere ($z \geq 500$km). A similar figure was shown in
\cite{2010ApJ...723L.185W}, their Fig. 4a, for a quiet Sun region observed
during the first flight of \sunrise{}.
For the investigations here, in an active region, we do not
find such a strong asymmetry in foot point strength as seen in
the quiet Sun. A substantial number of loops are close to the solid
line, which corresponds to equal strength of both foot points.
For the quiet Sun, symmetric or almost symmetric loops with a
leading foot point strength above $800$G have been absent.
This is different here and in active regions almost
symmetric loops exist even for foot point strengths of $2000$G and above.
As the scatter plot Fig. \ref{fig_mhs}d) and Table \ref{table1} show,
the majority of the active region loops has foot points with different
strength, but this effect is much less pronounced compared with
quiet Sun loops shown in \cite{2010ApJ...723L.185W} Fig. 4a).
\subsection{Plasma}
\begin{figure*}
\mbox{
\includegraphics[width=8.0cm, height=6.7cm]{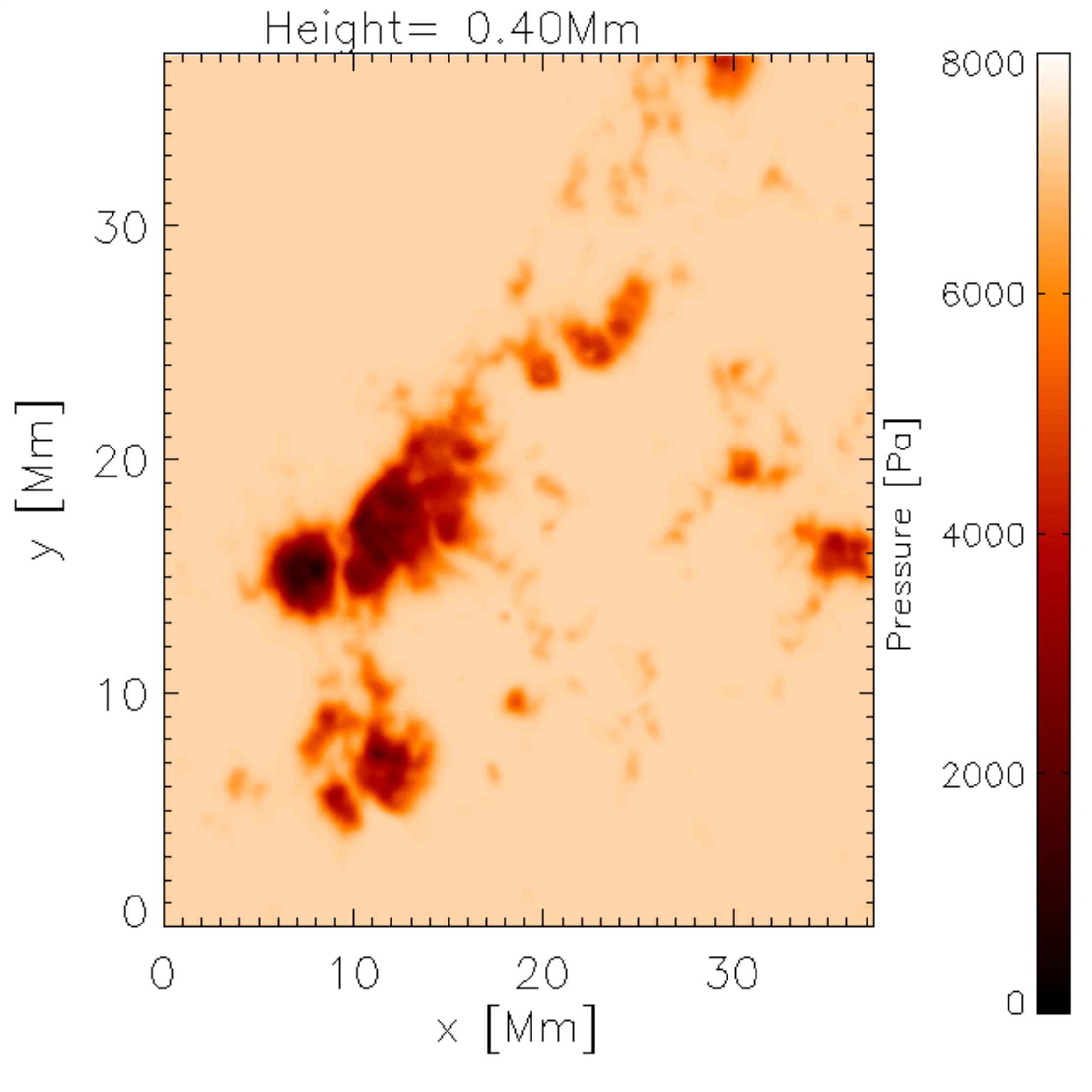}
\includegraphics[width=8.0cm, height=6.7cm]{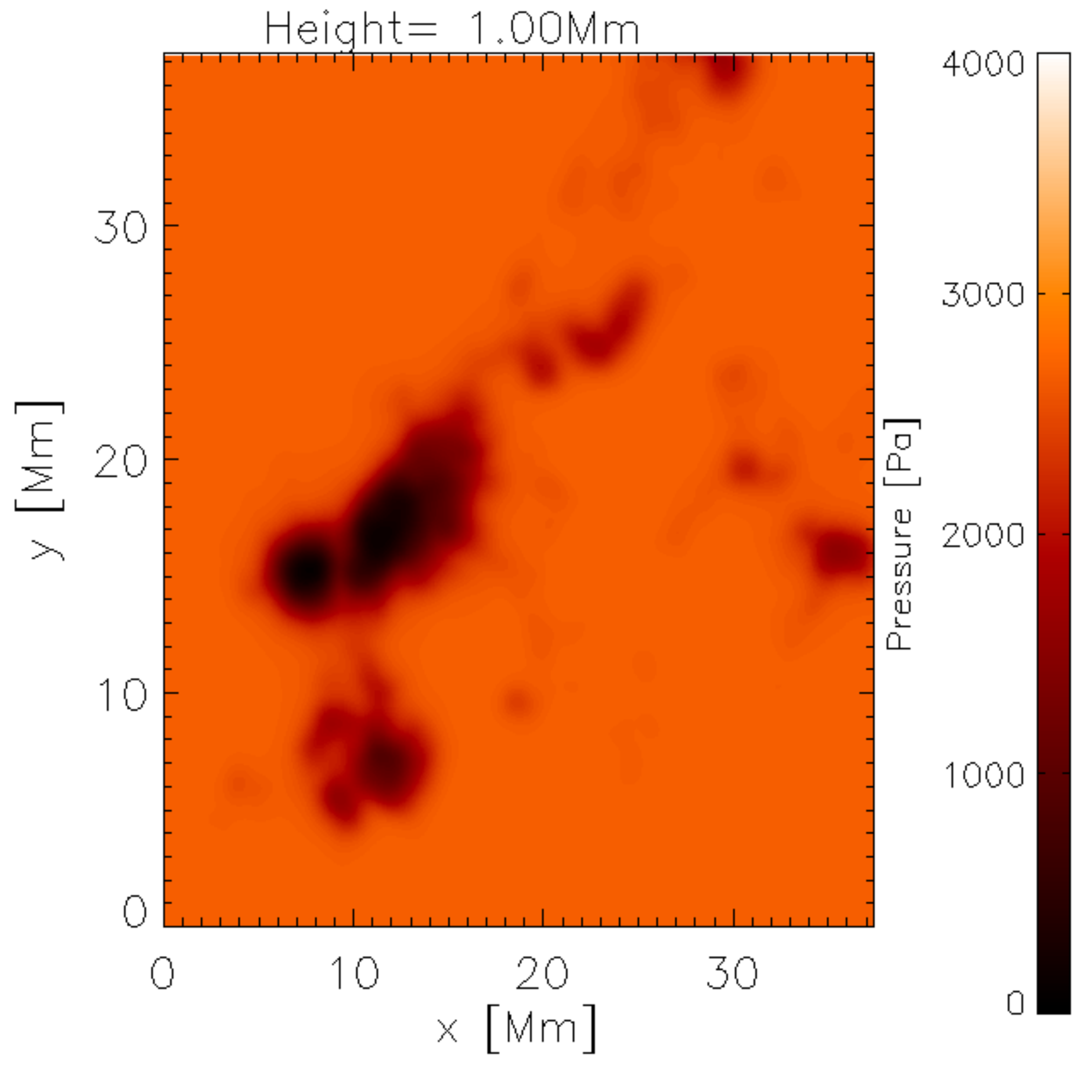}
}
\mbox{
\includegraphics[width=8.0cm, height=6.7cm]{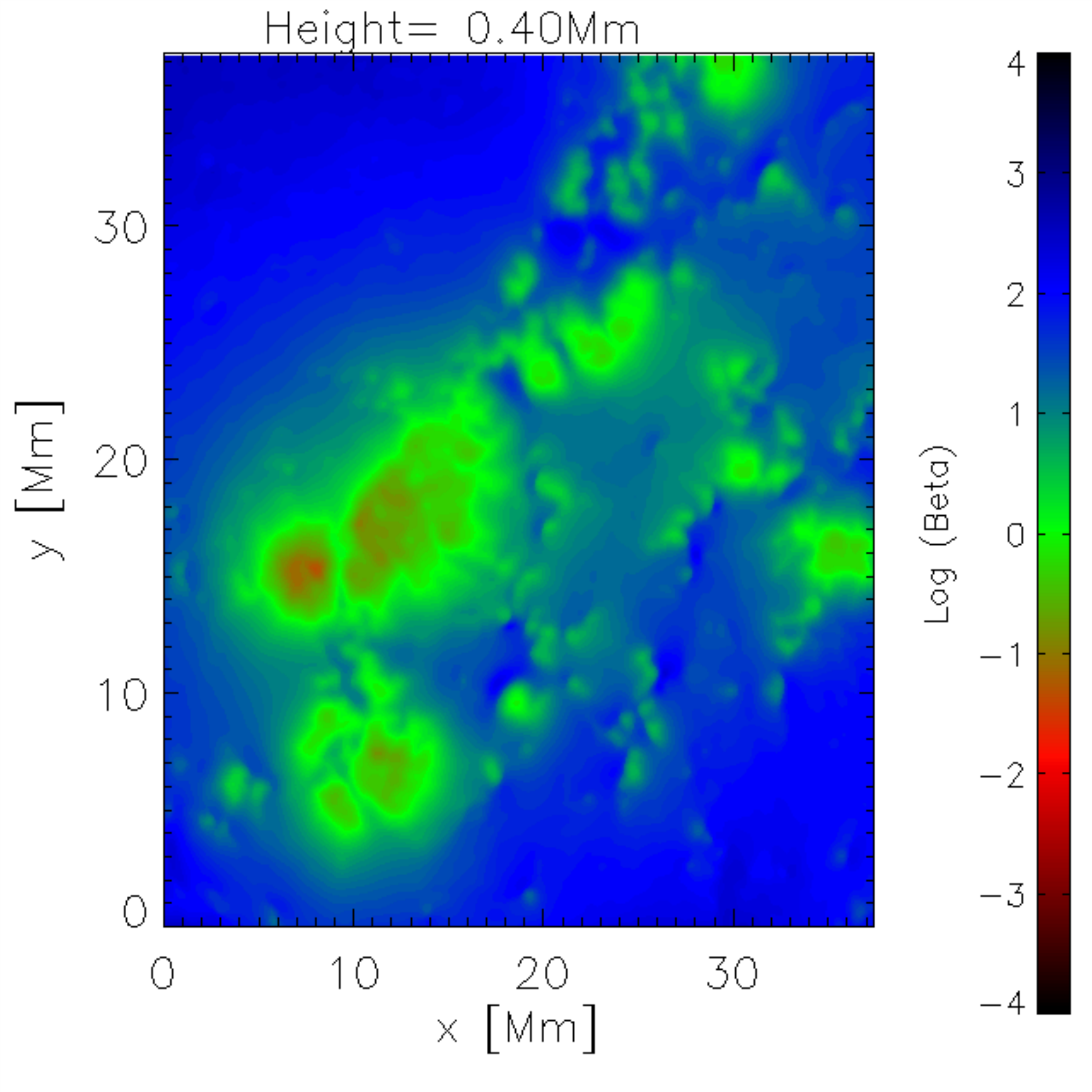}
\includegraphics[width=8.0cm, height=6.7cm]{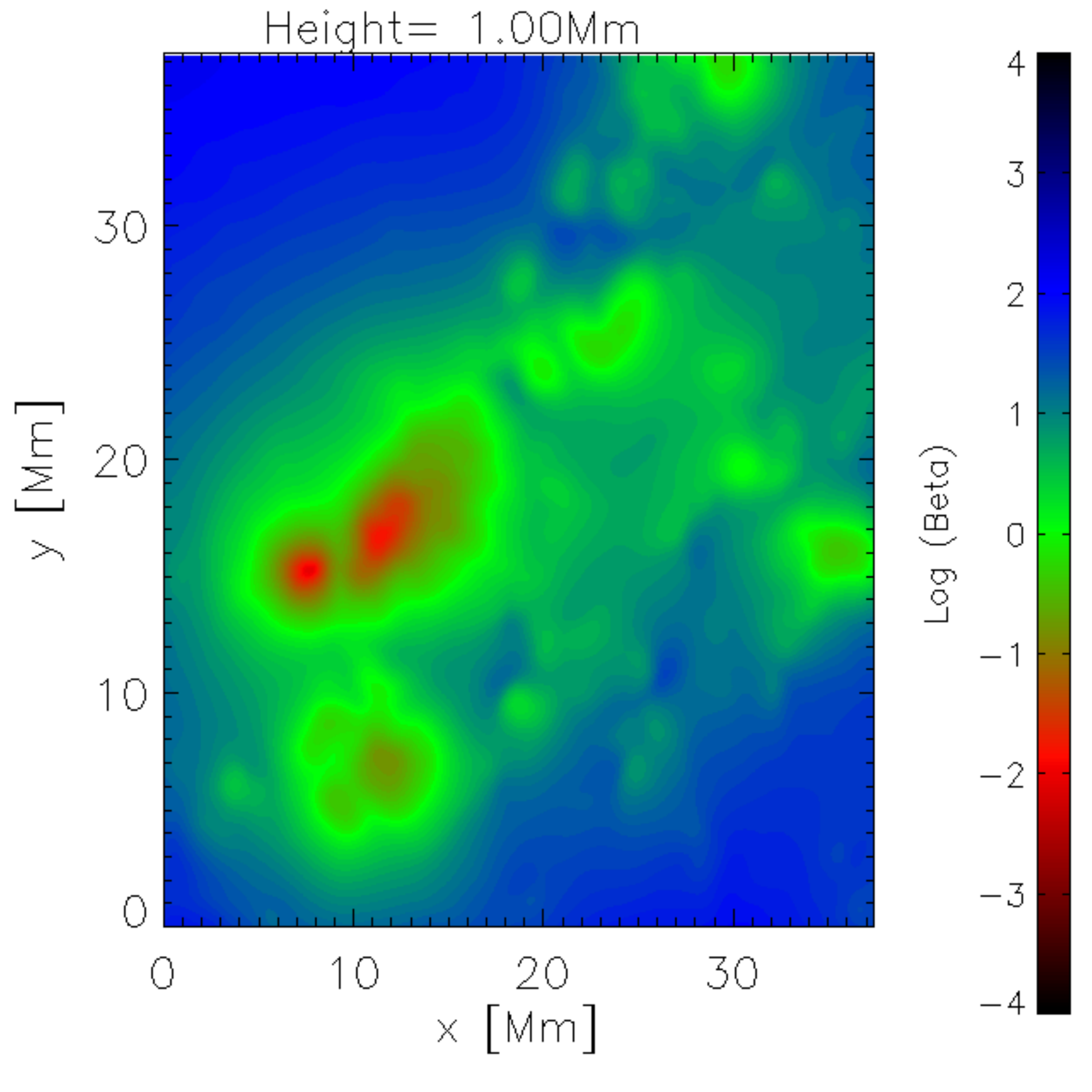}
}
\includegraphics[width=16.0cm, height=6.7cm]{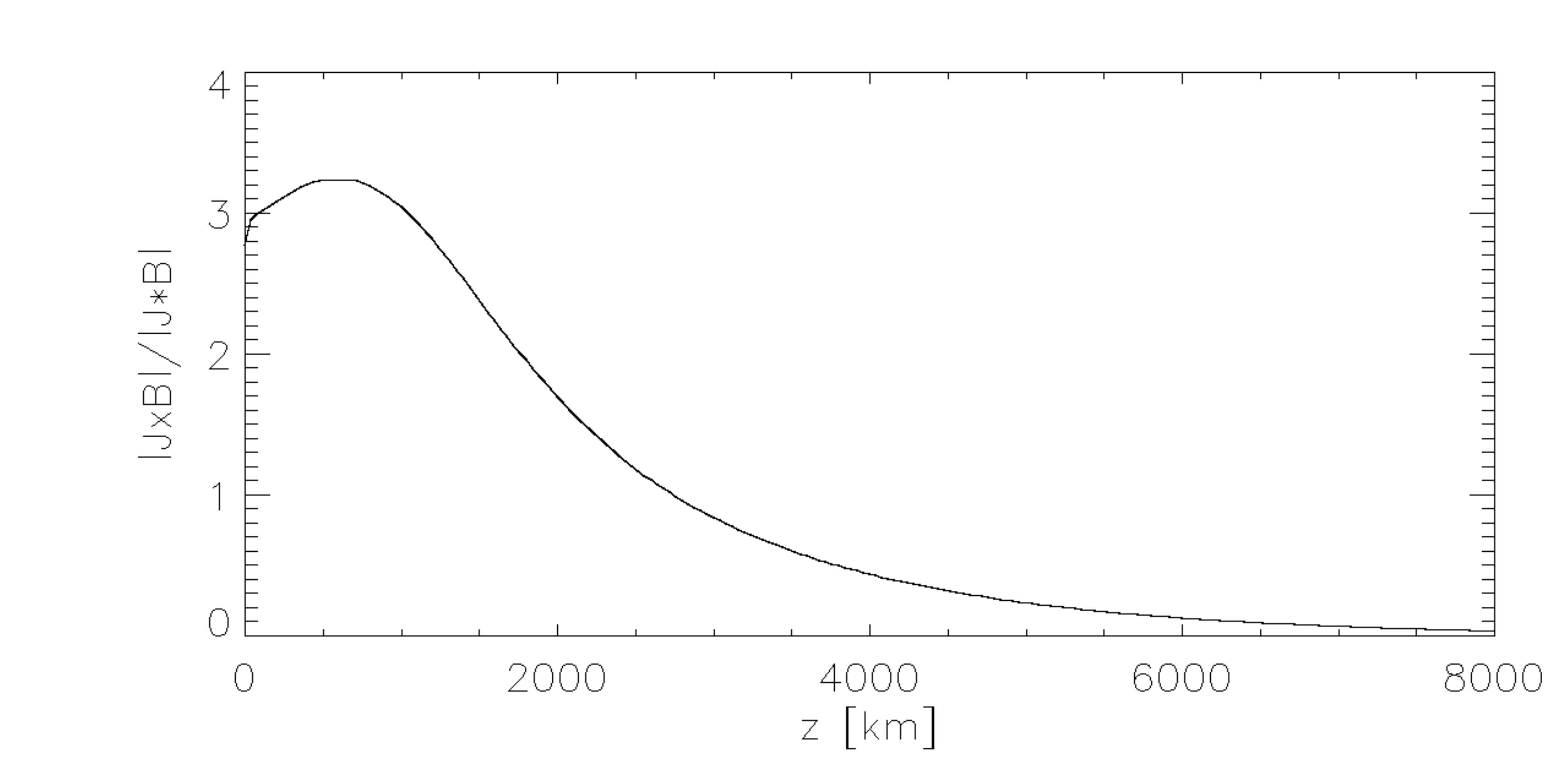}
\caption{Plasma pressure (top panels) and plasma $\beta$
(center panels) at height $400$ km and $1$ Mm for the first
snapshot at 23:39UT. The bottom panels shows the averaged Lorentz
force $\frac{|{\bf J} \times {\bf B}|}{|{\bf J} \cdot {\bf B}|}$
as a function of hight.}
\label{fig_plasma}
\end{figure*}
%\clearpage
%
Following {\it Paper I} the plasma pressure $p$ and density $\rho$
are divided into two parts, which are computed separately
and then added together.
The non-vanishing Lorentz force
is compensated by the gradient of the plasma pressure and in
the vertical direction also partly by the gravity force.
We compute the corresponding part of the plasma pressure
$p$ and density $\rho$ following the explanations given in {\it Paper I}.
Superimposed on this component a background plasma
(obeying a 1D-equilibrium of pressure gradient and gravity in the vertical
direction) is added to ensure a total positive density
and pressure. The top panels in Fig. \ref{fig_plasma} show
the plasma pressure in the upper photosphere (height $z=400$ km)
and mid chromosphere ($z=1$ Mm) for one snapshot from the beginning
of the time series. The center panels
in Fig. \ref{fig_plasma} show the plasma $\beta$ for the same heights.
The overall structure of these quantities, here shown for only one snapshot,
vary only very moderately in time.
A low plasma $\beta$ is a sufficient, but not necessary condition
for a magnetic field to be force-free.
To test whether our fields really are non-force-free,  we show the horizontal
averaged Lorentz force as a function of height in the bottom panel
of Fig. \ref{fig_plasma}. Shown is the dimensionless quantity
$\frac{|{\bf J} \times {\bf B}|}{|{\bf J} \cdot {\bf B}|}$, which compares
the importance of perpendicular and field aligned electric currents. The
quantity becomes zero for a vanishing Lorentz force. In the lower atmosphere
the perpendicular currents dominate (i.e.,
$|{\bf J} \times {\bf B}| > |{\bf J} \cdot {\bf B}|$)
 and there relative influence is maximum
at $z=600$km. Towards coronal heights, field aligned currents dominate
(i.e., $|{\bf J} \times {\bf B}| \ll |{\bf J} \cdot {\bf B}|$).

For the quiet-Sun region in
{\it Paper I} it was found that the used special class of magneto-static
equilibria are not flexible
enough to model the full FOV with a unique set of parameters $\alpha$
and $a$. The reason was that strongly localized magnetic elements in
an otherwise weak-field quiet-Sun area were incompatible with the
intrinsic linearity of the underlying equations. Similar limitations
do not occur, however, for the active region investigated in this
paper. The magnetic field of the large scale pore
(shown in dark blue in figure \ref{vecmag})
can be modelled significantly better with the linear approach than the
localized magnetic elements in {\it Paper I}. Furthermore, as explained in
section \ref{compute_alpha_a}, $\alpha$ and $a$ can be deduced
from measurements, which was not possible in the quiet Sun.
\subsection{Comparison with potential and force-free model}
\label{vergleich}
\begin{figure*}
\mbox{
\includegraphics[width=7.0cm, height=5.8cm]{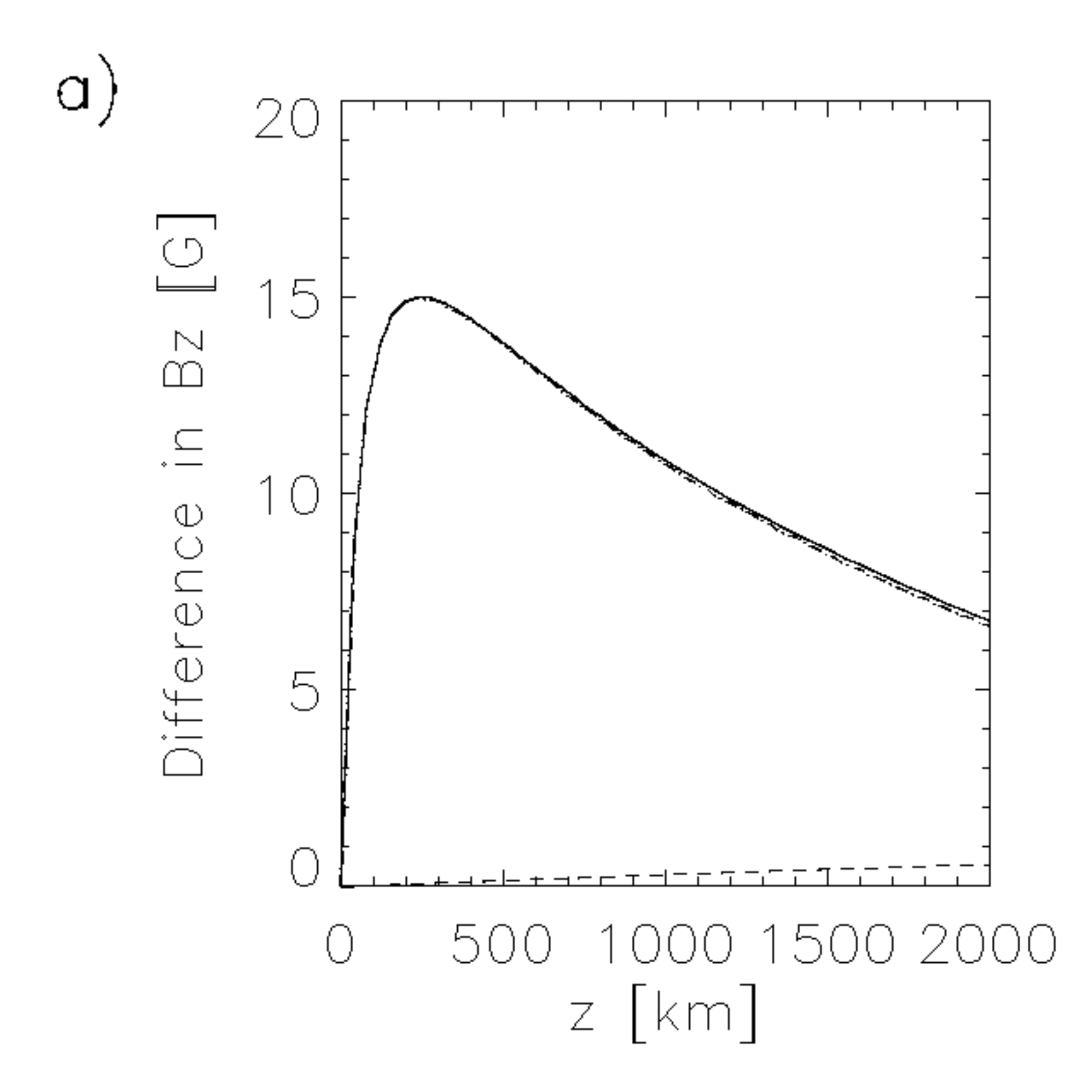}
\includegraphics[width=7.0cm, height=5.8cm]{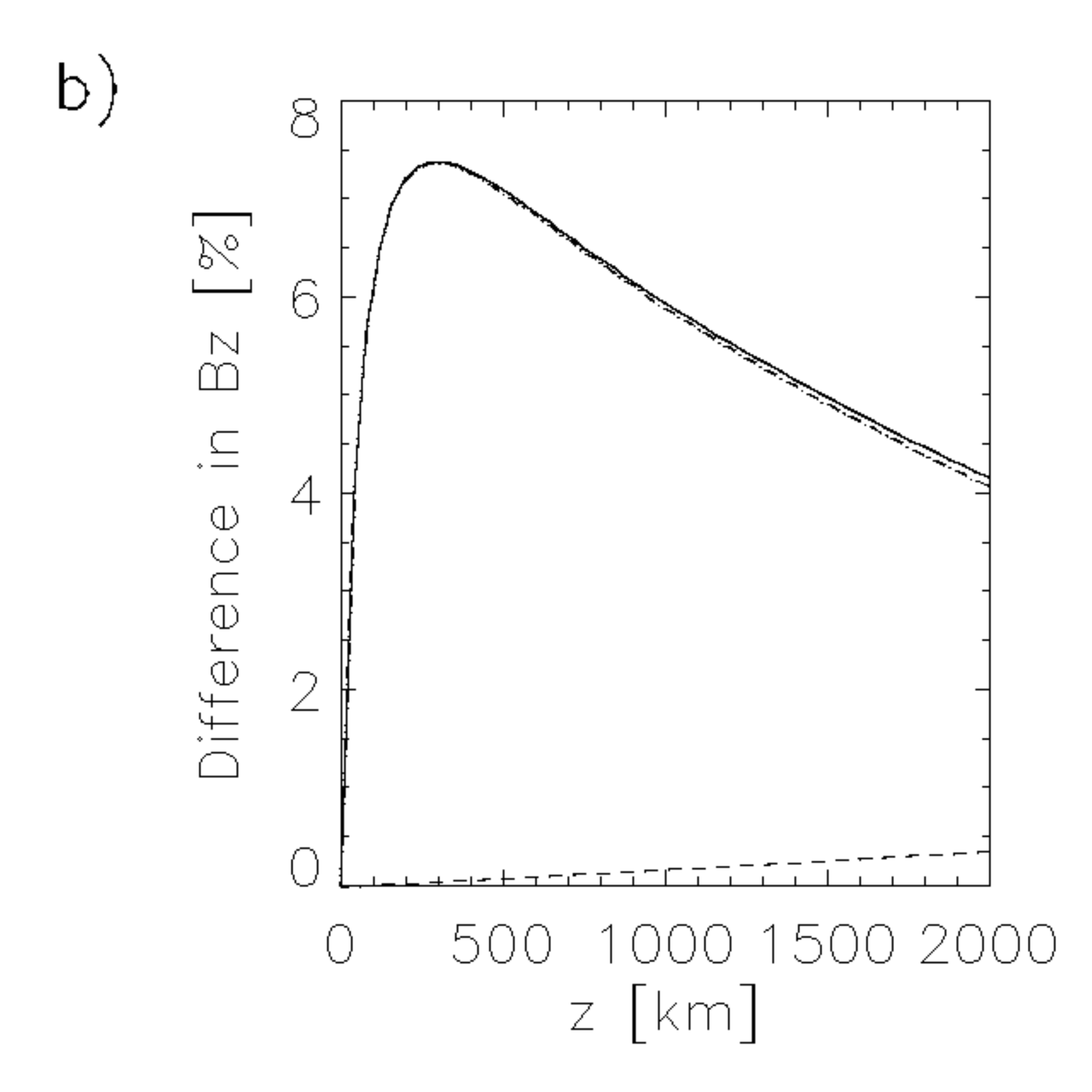}
}
\mbox{
\includegraphics[width=7.0cm, height=5.8cm]{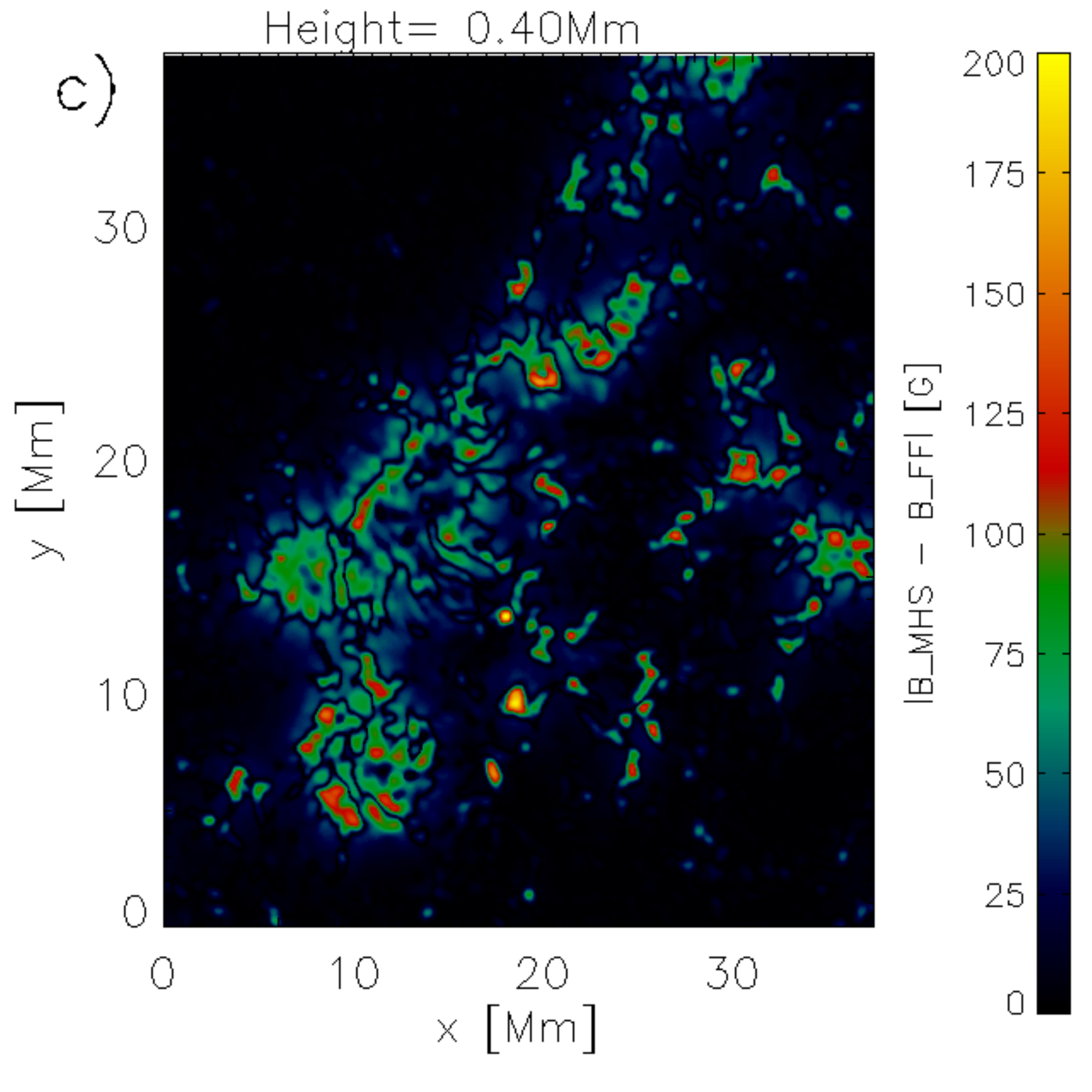}
\includegraphics[width=7.0cm, height=5.8cm]{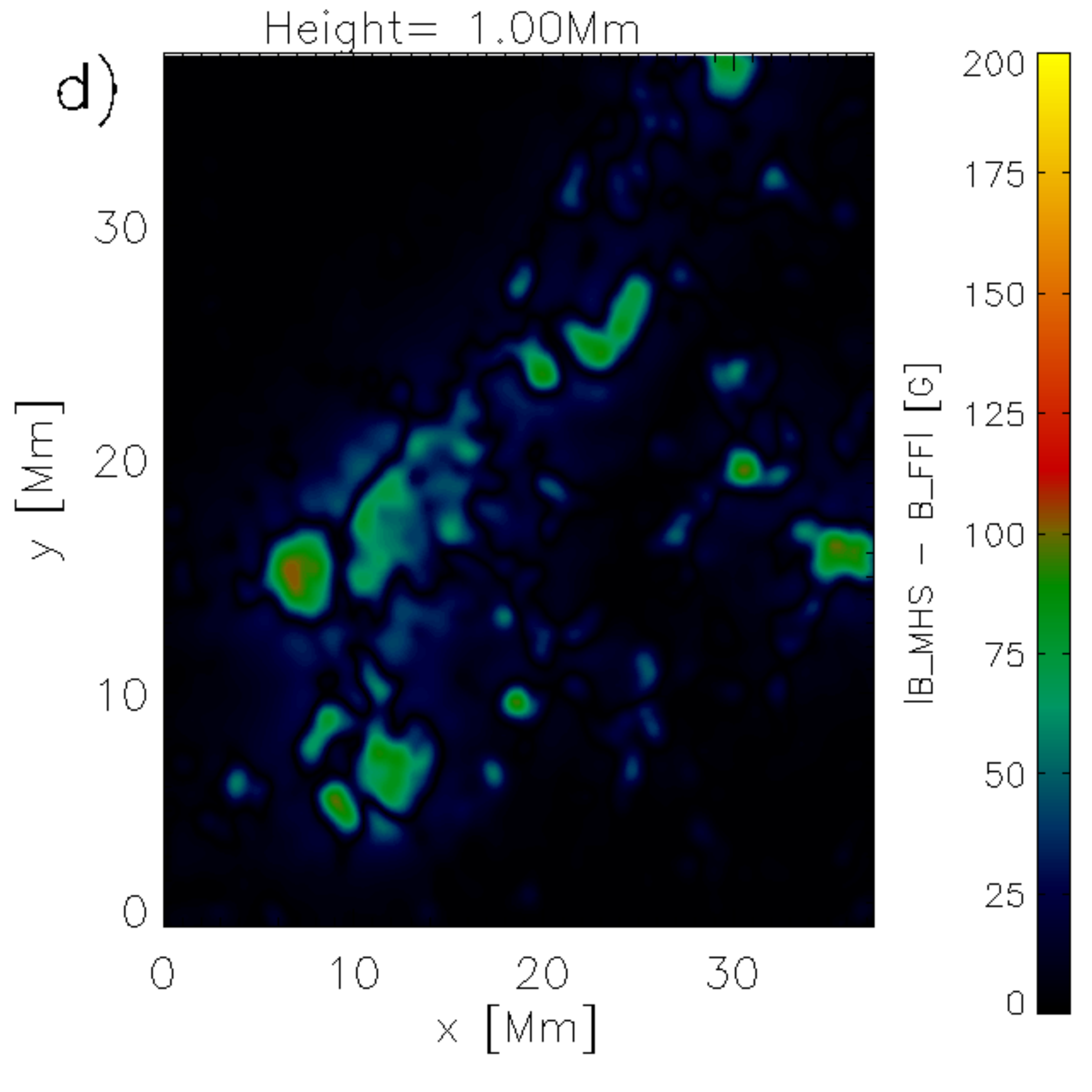}
}
\mbox{
\includegraphics[width=7.0cm, height=5.8cm]{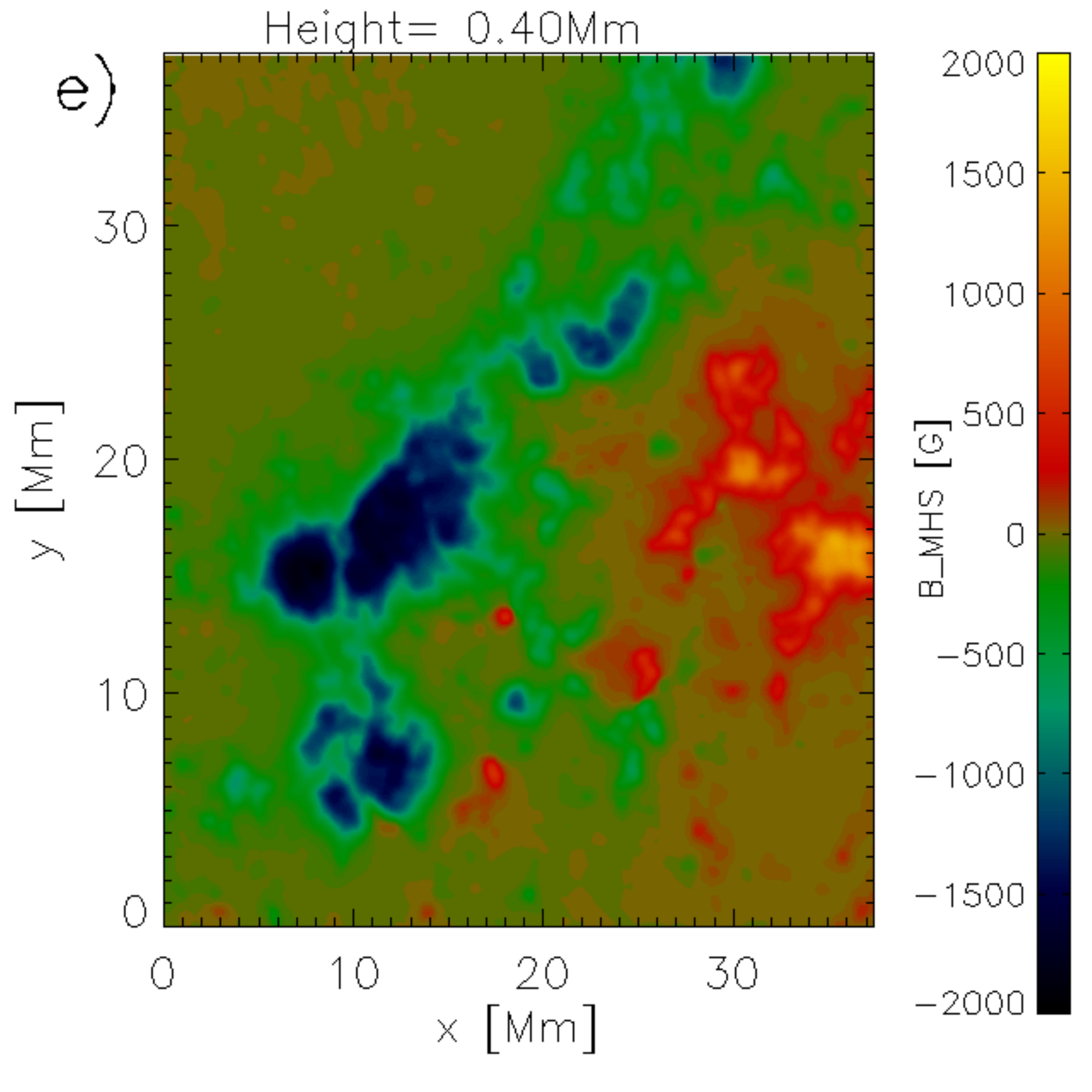}
\includegraphics[width=7.0cm, height=5.8cm]{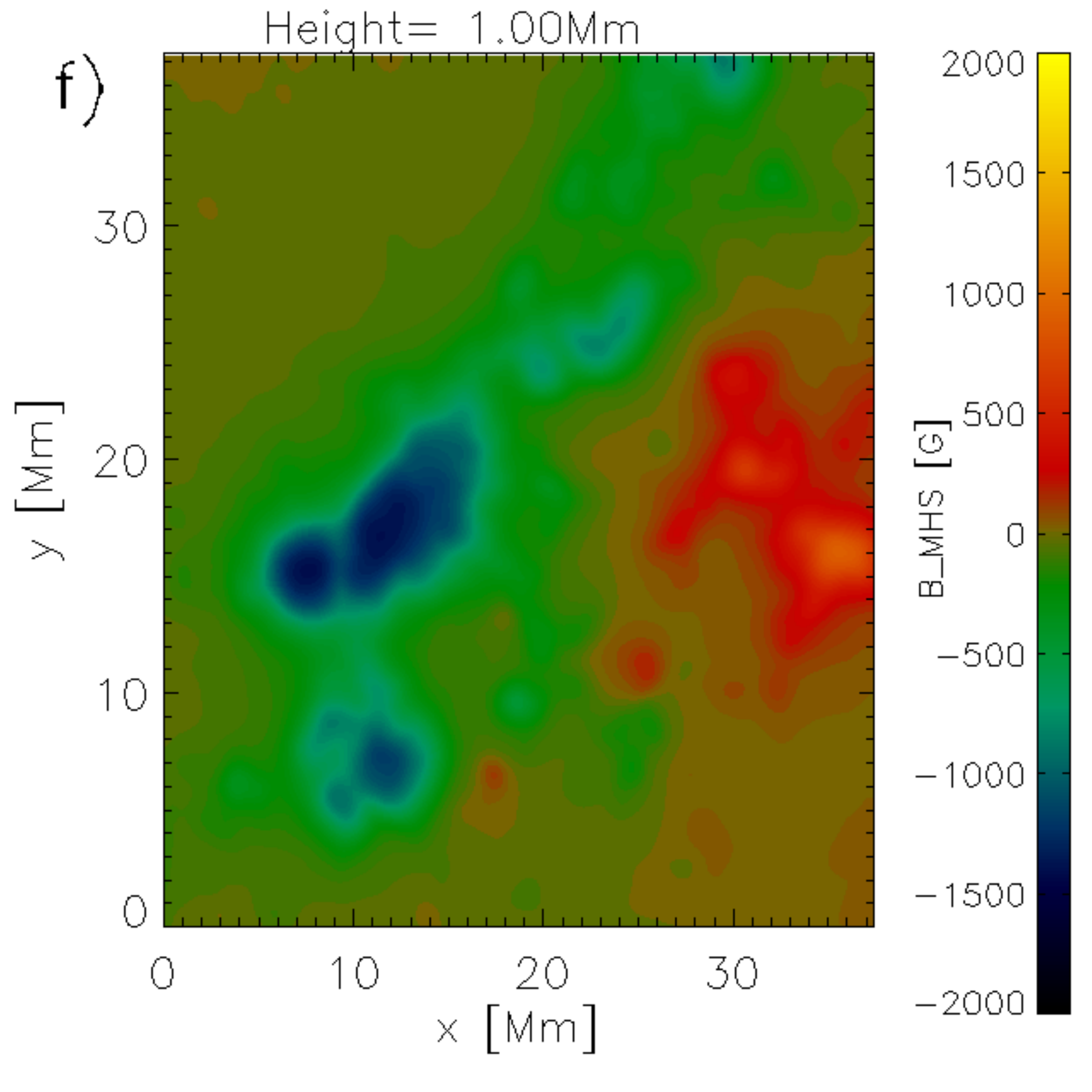}
}
\caption{a) Comparison of the vertical magnetic flux $z \rightarrow B_z(z)$
computed with different models. Average
difference in G of magneto-static (with $a=0.55$ and $\alpha L=-0.7$)
and linear-force-free (with $\alpha L=-0.7$ model (solid line),
magneto-static and potential field model (dash-dotted line) and
linear force-free and potential field model (dashed line).
b) Same as panel a), but the differences have been normalized with the
averaged absolute magnetic flux at every height $z$.
Panels c) and d) show the absolute differences of the vertical magnetic
field between a magneto-static and linear-force-free model in the
height $400$ km and $1$ Mm, respectively.
Panels e) and f) show $B_z$ of the MHS-model at the same heights.
All panels correspond to the first snapshot from IMaX at 23:39UT.
}
\label{fig_vergleich}
\end{figure*}
\begin{figure*}
\mbox{\includegraphics[width=7.0cm, height=5.8cm]{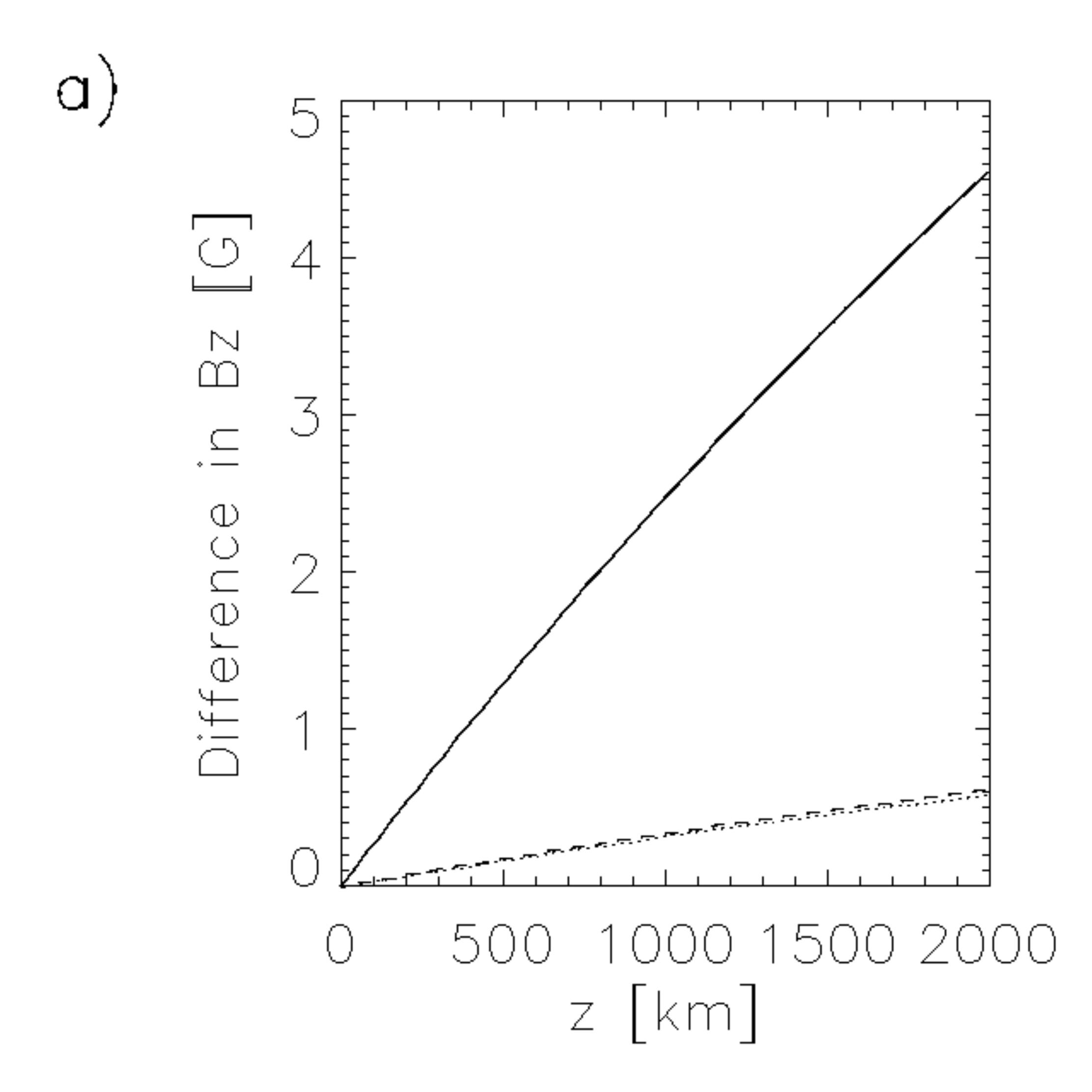}
\includegraphics[width=7.0cm, height=5.8cm]{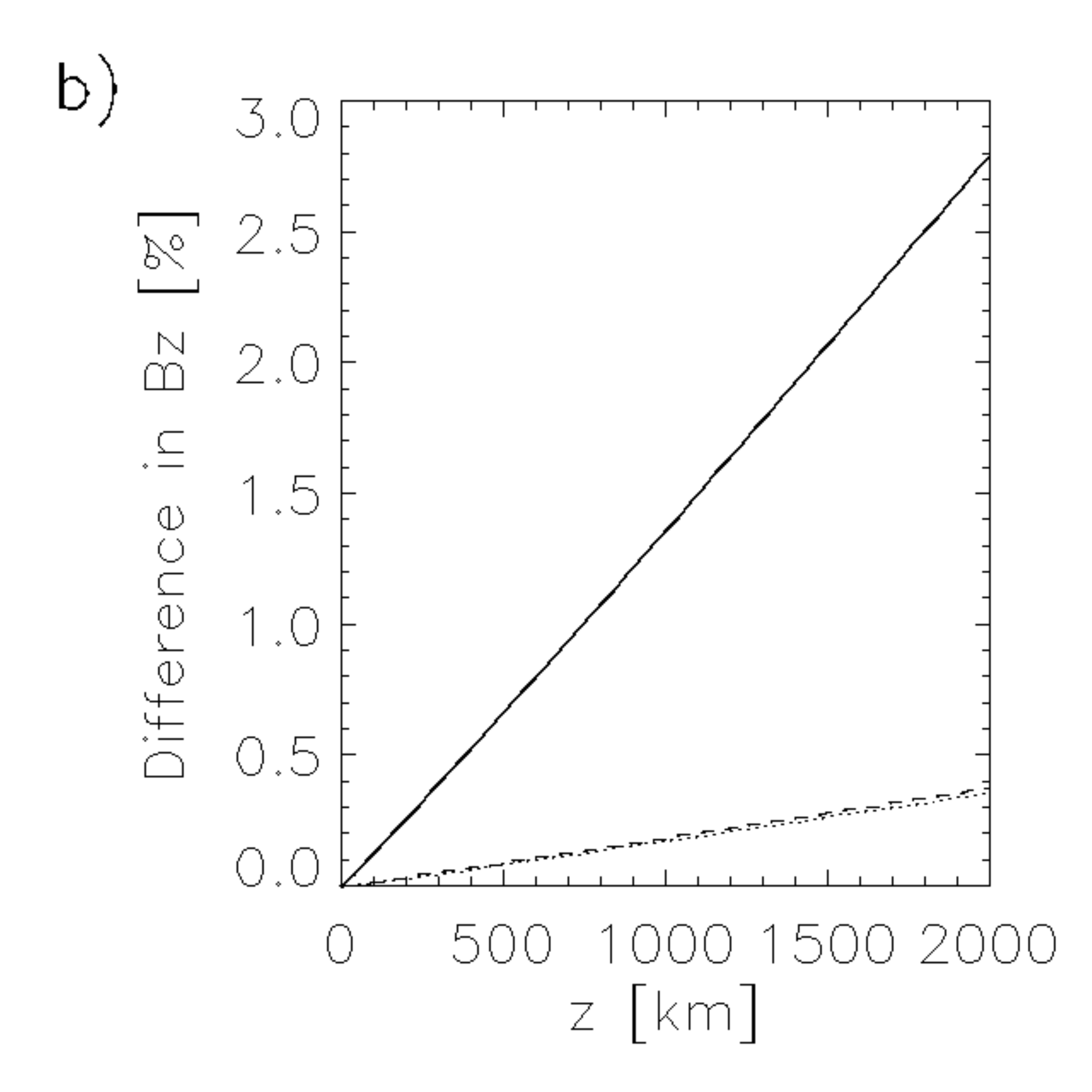}}
\mbox{\includegraphics[width=7.0cm, height=5.8cm]{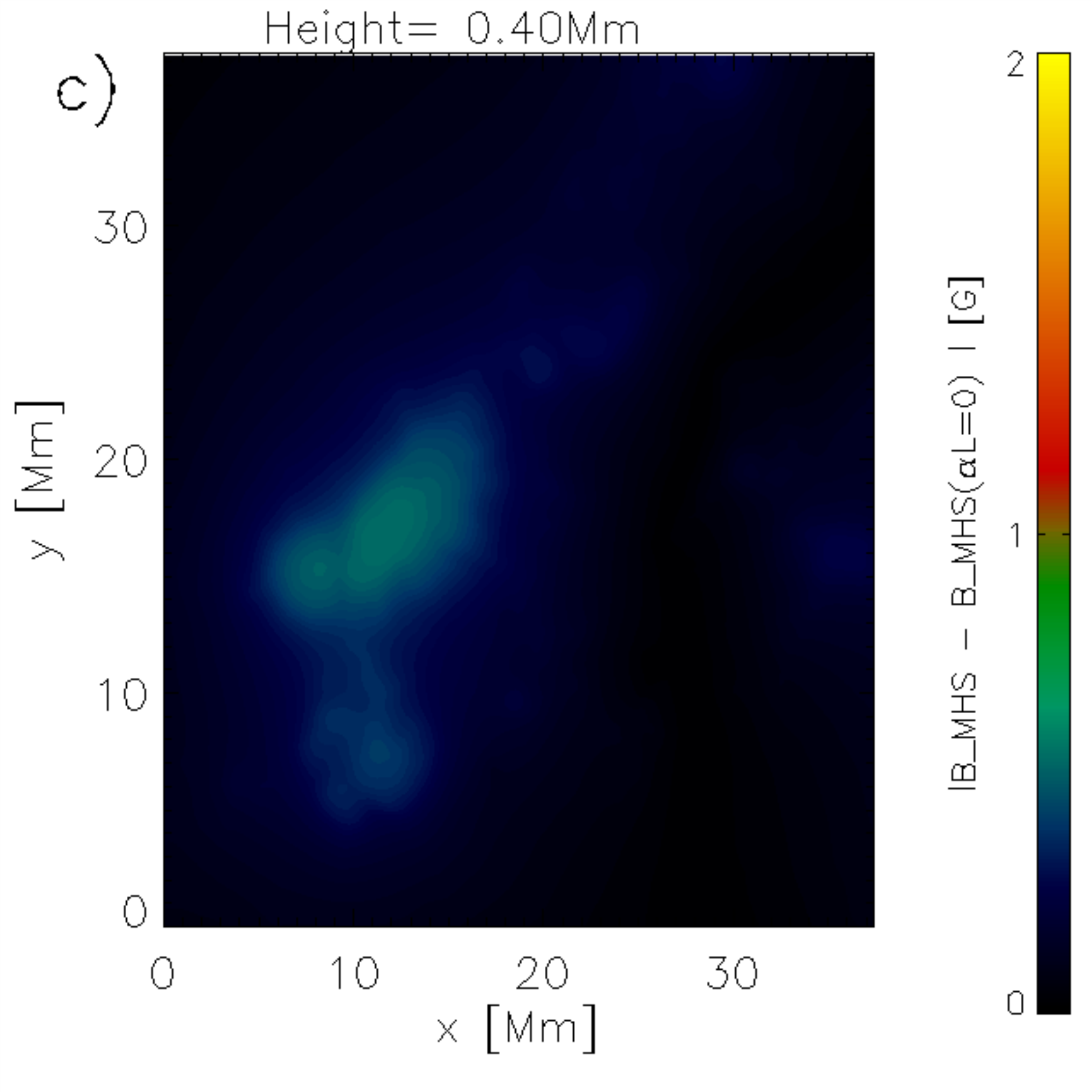}
\includegraphics[width=7.0cm, height=5.8cm]{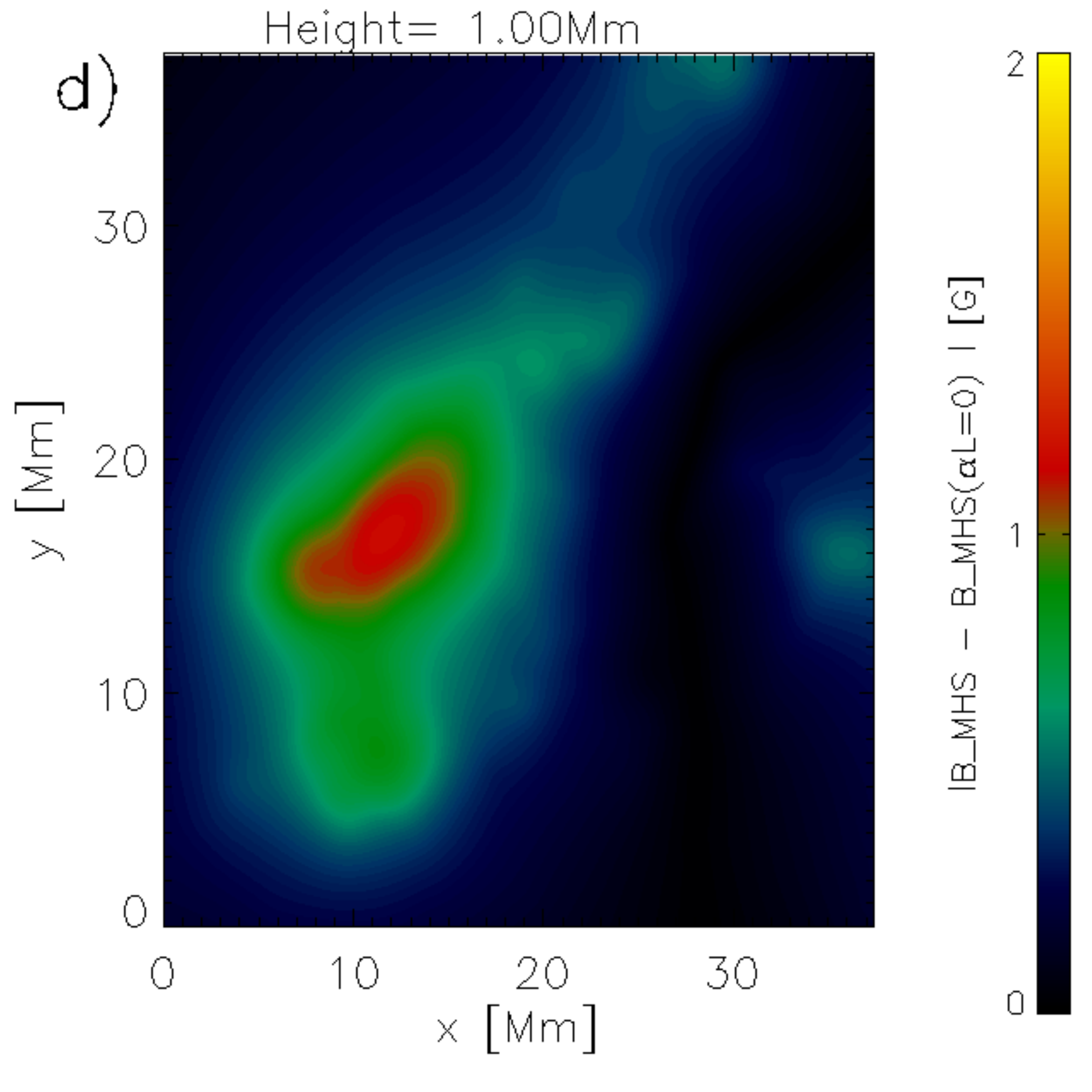}}
\mbox{\includegraphics[width=7.0cm, height=5.8cm]{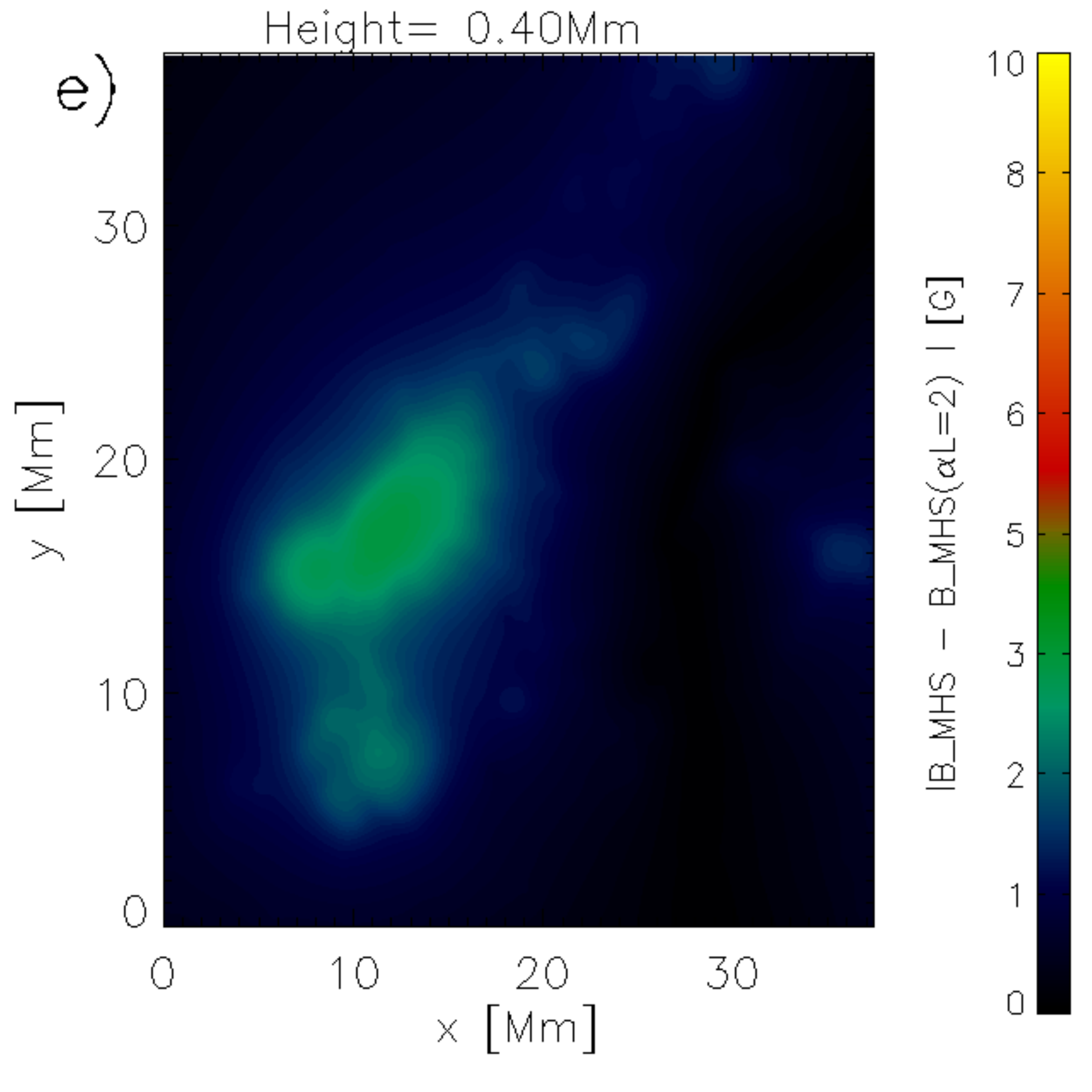}
\includegraphics[width=7.0cm, height=5.8cm]{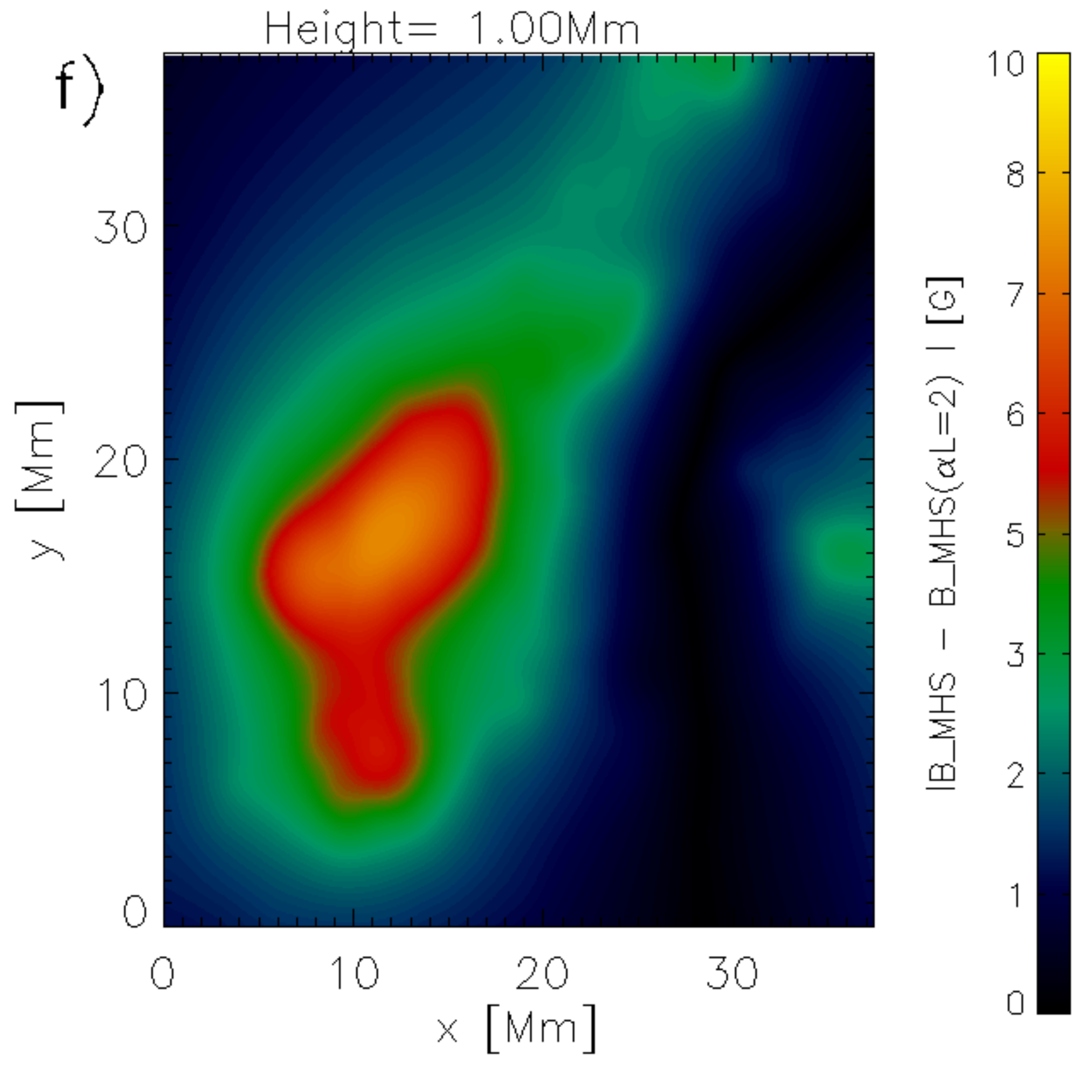}}
\caption{ a) Comparison of the vertical magnetic flux $z \rightarrow B_z(z)$
computed with magneto-static models  with different values of $\alpha$.
We show the average
difference in G of the original magneto-static model
$\alpha L =-0.7$  and a MHS-model with $\alpha L=0.0$ (dashed line)
and  $\alpha L=2.0$ (solid line).
b) Same as panel a), but the differences have been normalized with the
averaged absolute magnetic flux at every height $z$.
Panels c) and d) show the absolute differences of the vertical magnetic
field between the MHS-models with $\alpha L=-0.7$ and $\alpha L=0.0$ at the
height $400$ km and $1$ Mm, respectively.
Panels e) and f) show a comparison of
MHS-models with $\alpha L=-0.7$ and $\alpha L=2.0$.
Please note the different colour scales.
}
\label{fig8}
\end{figure*}
Simpler than magneto-static extrapolations are potential
and linear force-free models. Here we would like to point out
some differences. In Fig. \ref{fig_vergleich} panel a) we show
the average difference in vertical field strength $B_z$ as a function
of the height $z$ in the solar atmosphere.
The dashed line compares the result of a potential and linear force-free
model. As one can see the
differences are very small and increase linearly with height. This property
was already found for the quiet Sun in \cite{2010ApJ...723L.185W}.
The solid and dash-dotted lines compare the magneto-static model with
linear-force-free and potential fields, respectively. Both lines almost
coincide at low heights and differ only slightly higher. We find that
the magneto-static model deviates strongest from the other models at a height
$z= 240$ km. In panel b) the differences in $B_z$ have been normalized
with the (decreasing) average magnetic flux at every height. The curves
show, however, the same trend as for the absolute values, just the largest
difference is slightly shifted to $z= 280$ km. While this horizontal averaged
differences are with a maximum of about $15$ G and $7 \%$  relatively small,
the local deviation is significant, see
Fig. \ref{fig_vergleich} panels c) and d), which show the difference of
magneto-static and force-free field at a height of $z=400$ km and $z=1$ Mm,
respectively. It is not accidental that the differences in the vertical flux
between the magneto-static and the force-free model depict a somewhat similar
structure as the plasma pressure $p$ and the plasma $\beta$ shown in
Fig. \ref{fig_plasma}, because horizontal structures in the plasma
are the result of compensating a non-vanishing Lorentz-force.
The reason is that for strictly
force-free configurations the Lorentz force vanishes and consequently
the pressure gradient force has to be compensated by the gravity force alone.
Because the gravity force is only vertical in $z$, the pressure
cannot vary in the horizontal direction for force-free configurations.
Horizontal variations of the pressure, as shown in the top panels
of Fig. \ref{fig_plasma} occur in MHS-solutions, because the pressure
gradient force has to compensate the Lorentz force.
Consequently structures in the plasma occur in regions where force-free
and magneto-static
models differ most.  In panels e) and f) of Fig. \ref{fig_vergleich}
we show for comparison
the distribution of $B_z$ at the same heights. As one can see, the
maximum differences in $B_z$ are well below the maximum values of the
vertical field (by about a factor of ten). The largest differences are
in regions where $B_z$ is
strong and consequently the plasma pressure and plasma $\beta$ are low.
\subsection{Influence of the linear force-free parameter $\alpha$
on MHS equilibria.}
As one can see in Fig. \ref{time_a_alpha}, $\alpha L$ seems to
vary significantly in time and obtains values in the range
$\pm 2$. Here we would like to
investigate to which extend modifying the parameter $\alpha L$
affects the solution.
To do so, we compare (only  for the first snapshot) our original
linear magneto-static
solution with the deduced parameters $a=0.55$ and $\alpha L= -0.7$
with configurations, where  $\alpha L$ has been modified to
$\alpha L=0.0$ and $\alpha L=+2.0$, respectively.

 In Fig. \ref{fig8}a) we show
the average difference in vertical field strength $B_z$ as a function
of the height $z$ in the solar atmosphere.
Panel b) shows relative values normalized with averaged absolute
magnetic flux at every height $z$.
The solid (dotted) lines compare
the original MHS-equilibria  with the $\alpha L=+2.0$ ($\alpha L=0.0$)
ones, respectively. Naturally a larger discrepancy of $\alpha$ results
in larger difference of the resulting fields. The influence is, however,
much smaller than the comparison of MHS-equilibria and potential and linear
force-free fields shown in Fig. \ref{fig_vergleich}. A major difference
is, however, that the discrepancy of MHS-solutions with different values
of $\alpha$ increase with height. Such a property is well known already
from the comparison of potential and linear force-free fields in
\cite{2010ApJ...723L.185W}. In the top panels of Fig. \ref{fig8}(a and b)
we overplot again
the difference of a linear force-free field with
$\alpha L= -0.7$ and a potential field with $\alpha L=0.0$ with dotted
lines. (This quantity was shown already in Fig. \ref{fig_vergleich} with
a different axis scale). The dotted and dashed line almost coincide
(about $5\%$ difference) and we can conclude that modifying $\alpha$
in MHS-equilibria has a a similar effect as in linear force-free
configurations. Fig. \ref{fig8} panels c-f show the differences of
the MHS-solutions in the height $400$km and $1$Mm, respectively.
These images should be compared with the corresponding panels in
Fig. \ref{fig_vergleich}, but please note the very different colour scales
(by a factor of 100 between panels c,d in Figs. \ref{fig_vergleich} and
\ref{fig8} and by a factor of 20 between panels c,d in
Figs. \ref{fig_vergleich} and panel e,f in Fig. \ref{fig8}).
For low-lying structures
the influence of changing $\alpha$ is therefor very small. Far more
important for the structure in photospheric and chromospheric heights
 is the force parameter $a$.

\section{Discussion and outlook}
\label{sec:outlook}
Within this work we applied a special class of magneto-static equilibria
to model the solar atmosphere above an active region. As boundary
condition we used measurements of the photospheric magnetic field
vector obtained by \sunrise{}/IMaX, which have been embedded into SDO/HMI
active-region magnetograms. The used approach models the
3D magnetic field in the solar atmosphere self-consistently with the
plasma pressure and density.

Pressure gradient and gravity forces are
important only in a relatively thin (about 2 Mm) layer containing the
photosphere and chromosphere. Thanks to the high spatial resolution of
IMaX, we were able to resolve this non-force-free layer with $50$
grid points. In {\it Paper I} we discussed the limitations of applying our
model to the quiet Sun, where strong localized flux elements make
a linear approach less favourable. While, in principle,
nonlinear models are generically more flexible,
we found the active region investigated here to be far more suitable
for a linear model than the quiet Sun.
In particular, the entire domain could be modelled without running
into problems with negative densities and pressures that plagued
the application to the quiet Sun.
We derived free model parameters from IMaX and a unique set of
these parameters was used in the entire modelling domain.  The
parameter $\alpha$ controls field aligned currents and the parameter
 $a$ horizontal currents. While the currents controlled by $a$ are
strictly horizontal, they have a field line parallel part
and a part perpendicular to the field. The latter one is responsible for the
finite Lorentz-force and deviation from force-freeness.

Nevertheless,
a linear magneto-static model can only be a lowest order approximation.
Shortcomings of any linear approach is that the generic non-linear
nature of most physical systems is not taken into account. For equilibria
in the solar atmosphere this means that strong current concentrations and
derived quantities like the spatial distribution of $\alpha(x,y)$ are
not modelled adequately.
The situation is somewhat similar to the history of force-free coronal
models, where linear force-free active region models have been routinely used
\citep[see,e.g.][]{1978SoPh...58..215S,1989ApJS...69..323G,1992A&A...258..535D,
1992A&A...257..278D,2002SoPh..208..233W,2004A&A...428..629M,
2005Sci...308..519T} before nonlinear force-free models entered the scene.
Nonlinear magneto-static extrapolation codes have been developed and tested
with synthetic data in
\cite{wiegelmann:etal06,wiegelmann:etal07,2016SoPh..tmp..182G}. While,
in principle, it is straight forward to apply these models to data
from \sunrise{}/IMaX, the implementation details are still challenging,
just as this was
the case about a decade ago for nonlinear force-free models.
A number of problems still need to be solved before nonlinear magneto-static
equilibria can be reliably calculated and such models can be routinely
applied to solar data.
Two issues remain open and are to be dealt with for applying nonlinear
magneto-static models to data: i) the noise in photospheric
magnetograms, and ii) the problem that the plasma  $\beta$ varies over
orders of magnitude within the computational volume, which slows down
the convergence rate of such codes \citep[see][for details]{wiegelmann:etal06}.
That magneto-static codes are slower than corresponding force-free
approaches has been reported also recently in \cite{2016SoPh..tmp..182G}
for a Grad-Rubin like method.

Linear magneto-static
equilibria, as computed in this work, can serve as initial
conditions for nonlinear computations. Last but not least, one
should understand the  transition from magneto-static to
force-free models above the mid chromosphere. While, in principle,
the magneto-static approach includes the force-free one automatically
for $\beta \rightarrow 0$, the computational overhead of computing
magneto-static equilibria in low $\beta$ regions  is severe. In
low $\beta$ regions, force-free codes can be applied because
the back-reaction of the plasma onto the magnetic field is small and
the numerical convergence is faster.

In this paper we applied  a linear magneto-static model
to compute the magnetic field in the solar atmosphere above
an active region. We modelled
the mixed $\beta$  layer of photosphere
and chromosphere, which required high resolution photospheric
field measurements as boundary condition.
This work is the second part of applying a linear magneto-static model to
high resolution photospheric measurements. In {\it Paper I} the model
was applied to the quiet Sun. The quiet Sun
is composed of small, concentrated (strong) magnetic
elements and large inter-net-work regions with weak magnetic field
in the photosphere. This property
is a challenge to the linear magneto-static model, because the plasma pressure
disturbances caused by strong and strongly localized magnetic elements,
require a background pressure, which results in an unrealistic high
average plasma $\beta$. As pointed out in {\it Paper I} the model
can be applied locally around magnetic elements, but does not permit
a meaningful modelling of large quiet-Sun areas containing magnetic
elements of very different strengthes. Strong localization of magnetic
elements and the linearity of the model are a contradiction. In active
regions, large magnetic pores and sunspots dominate the magnetic
configuration. The wider coverage by  strong fields
in active regions is more consistent with the limitations of a linear
model. Furthermore the free model parameters $\alpha$ and $a$ can
be deduced from horizontal magnetic field measurements in active regions,
which was not possible in the quiet Sun, due to the poor signal-to-noise
ratio.
%
%\clearpage

\section*{Acknowledgements}
\begin{acknowledgements}
The German contribution to \sunrise{} and its reflight was funded by the
Max Planck Foundation, the Strategic Innovations Fund of the President of the
Max Planck Society (MPG), DLR, and private donations by supporting members of
the Max Planck Society, which is gratefully acknowledged. The Spanish
contribution was funded by the Ministerio de Econom\'ia y Competitividad under
Projects ESP2013-47349-C6 and ESP2014-56169-C6, partially using European FEDER
funds. The HAO contribution was partly funded through NASA grant number
NNX13AE95G. This work was partly supported by the BK21 plus program through
the National Research Foundation (NRF) funded by the Ministry of Education of
Korea.
The used HMI-data are courtesy of NASA/SDO and the HMI science team.
TW acknowledges DLR-grant 50 OC 1301 and DFG-grant WI 3211/4-1.
TN acknowledges support by the UK's Science and Technology Facilities
Council via Consolidated Grants ST/K000950/1 and ST/N000609/1.
DN was supported from GA\,\v{C}R under grant numbers
16-05011S  % Dieter
and
16-13277S. % Petr Jelinek
The Astronomical Institute Ond\v{r}ejov is supported by the project
RVO:67985815.
The National Solar Observatory (NSO) is operated by the Association of
Universities for Research in Astronomy (AURA) Inc. under a cooperative
agreement with the National Science Foundation.
\end{acknowledgements}
%

%\bibliographystyle{aa}
%\bibliographystyle{plain}
%\bibliographystyle{plainnat}
%\bibliographystyle{abbrvnat}
%\bibliographystyle{unsrtnat}
%\bibliographystyle{alpha}
%\bibliographystyle{named}
%\bibliography{mhs}
\pagebreak
\clearpage
%\appendix

\end{document}